\documentclass[times,final]{elsarticle}
\usepackage{graphicx} 

\usepackage[utf8]{inputenc}
\usepackage[T1]{fontenc}
\usepackage{amsmath, amsfonts, amssymb}
\usepackage{wrapfig}
\usepackage{natbib}
\usepackage{url}
\usepackage{hyperref}
\usepackage{eso-pic}
\usepackage{amssymb}
\usepackage{epstopdf}
\usepackage{chngcntr}
\usepackage{subcaption}
\usepackage{placeins}

\usepackage{amssymb}

\usepackage{amsmath}
\usepackage{accents}
\usepackage{graphicx}
\usepackage[colorinlistoftodos]{todonotes}
\usepackage{lipsum} 
\usepackage{floatrow} 
\usepackage[hypcap=false]
{caption} 
\usepackage{subcaption}

\usepackage{listings}
\usepackage{url}
\usepackage{courier}
\usepackage{wrapfig}
\usepackage{color,soul}
\definecolor{codegreen}{rgb}{0,0.6,0}
\usepackage{chngpage}
\usepackage{xcolor}
\usepackage{bm}
\usepackage{amsmath,amssymb}
\usepackage{verbatim}
\newlength{\mycolwidth}
\usepackage{array}
\newcolumntype{Z}{>{$}p{\mycolwidth}<{$}}
\usepackage{algorithmicx}
\usepackage{algorithm} 
\usepackage{refcount}
\usepackage{algpseudocode} 
\usepackage{placeins} 
\MakeRobust{\Call}

\usepackage{amsfonts}
\usepackage{graphicx}
\usepackage{epstopdf}
\usepackage{amsmath} 

\usepackage{lscape}
\usepackage{booktabs}

\usepackage{tikz,lipsum,lmodern}
\usepackage[most]{tcolorbox}

\usepackage{pifont}

\DeclareMathAccent{\svec}{\mathord}{letters}{126}
\newcommand\acclrvec[1]{\accentset{\,\leftrightarrow}{#1}}
\newcommand\stvec[1]{\mathbf #1}
		
\newcommand\ssvec[1]{\acclrvec{\stvec{#1}}}	

\newcommand\cssvec[1]{\acclrvec{\tilde{\stvec{#1}}}}

\usepackage{scalerel,stackengine}
\stackMath

\newcounter{bla}

\usepackage{lineno}

\counterwithin{figure}{section}
\title{A comparison of h- and p-refinement to capture wind turbine wakes}

\author[1]{Hatem Kessasra\corref{cor}}
\author[1,2]{Marta Cordero-Gracia}
\author[1,2]{Mariola Gómez}
\author[1,2]{Eusebio Valero}
\author[1,2]{Gonzalo Rubio}
\author[1,2]{Esteban Ferrer}
\address[1]{ETSIAE-UPM - School of Aeronautics, Universidad Politécnica de Madrid, Plaza Cardenal Cisneros 3, E-28040 Madrid, Spain}
\address[2]{Center for Computational Simulation, Universidad Politécnica de Madrid, Campus de Montegancedo, Boadilla del Monte, 28660 Madrid, Spain}

\cortext[cor]{Corresponding author. \textit{Email:} hatem.kessasra@alumnos.upm.es}

\begin{document}

\begin{abstract}
This paper investigates a critical aspect of wind energy research - the development of wind turbine wake and its significant impact on wind farm efficiency. The study focuses on the exploration and comparison of two mesh refinement strategies, h- and p-refinement, in their ability to accurately compute the development of wind turbine wake. The h-refinement method refines the mesh by reducing the size of the elements, while the p-refinement method increases the polynomial degree of the elements, potentially reducing the error exponentially for smooth flows. A comprehensive comparison of these methods is presented that evaluates their effectiveness, computational efficiency, and suitability for various scenarios in wind energy. The findings of this research could potentially guide future studies and applications in wind turbine wake modeling, thus contributing to the optimization of wind farms using high-order h/p methods. This study fills a gap in the literature by thoroughly investigating the application of these methods in the context of wind turbine wake development.
\end{abstract}


\maketitle

\tableofcontents


\section{Introduction}

High-order numerical methods have garnered significant attention in computational fluid dynamics due to their ability to achieve high accuracy with fewer computational resources. These methods, which utilize polynomial approximations of the solution, offer a promising alternative to traditional low-order techniques, particularly in complex flow simulations where precision is paramount. 
Low-order numerical techniques (most commercial codes) suffer from non-negligible numerical errors (e.g., dissipative and dispersive errors) that can increase unphysical dissipation and dispersion of flow structures and provide unrealistic results \cite{Juan_vonNeumann,Manzanero_2020}.
In contrast, high-order methods offer a solution to these challenges by allowing greater accuracy with fewer degrees of freedom ($DOF$), particularly when using mesh refinement techniques. High-order h/p methods are characterized by their ability to use mesh refinement through an increased number of mesh nodes (h-refinement) and/or polynomial enrichment (p-refinement) to achieve highly accurate solutions. Such high-order polynomial methods produce an exponential decay of the error for sufficiently smooth solutions instead of the algebraic decay characteristic of low-order techniques \cite{WorkshopDG}.

This paper aims to explore and compare two refinement strategies within high-order methods, h- and p-refinement, focusing on their efficacy in capturing intricate dynamics such as wind turbine wake development. The h-refinement method, which involves refining the mesh by reducing element sizes, is effective in resolving fine details but increases computational complexity due to the necessity of hanging nodes. Conversely, the p-refinement method, which increases the degree of polynomial within each element, promises exponential error reduction for smooth flows without the need for hanging nodes, potentially simplifying the mesh structure and reducing computational costs. 

In the context of wind energy, the accurate modeling of wakes after wind turbines is critical, as these simulations directly influence the optimization and efficiency of wind farm layouts \cite{annurev:/content/journals/10.1146/annurev-fluid-010816-060206}.
One of the most challenging aspects of wind farm operations is the development and impact of wind turbine wakes. The wake effect, characterized by reduced flow velocities and increased turbulence downstream of a turbine, can lead to significant energy losses when neighboring turbines are affected. Understanding and accurately simulating these wakes is essential for optimizing turbine placement and overall wind farm design, thereby enhancing power output and reducing maintenance costs.
%
%
Accurate simulations of wakes have significant implications:
\begin{itemize}
  \item  Wake Effect on Power Performance: Turbines operating in the wake of upstream turbines experience lower wind speeds and higher levels of turbulence, leading to a decrease in power output and potential damage due to fatigue loads. Accurate wake simulations allow for a better understanding of these effects, enabling more effective turbine placement and operation strategies.


 \item Implications for Wind Farm Design: The knowledge gained from wake simulations can significantly influence wind farm design. By strategically positioning turbines to minimize wake interference, it is possible to enhance the overall power performance of the wind farm. This approach also helps in reducing maintenance costs and extending the lifespan of the turbines.

\end{itemize}

A wind turbine simulation can easily involve 50 million degrees of freedom ($DOF$) when using low-order methods and actuator lines. Using high-order methods, the total number of $DOF$ can be significantly reduced while maintaining the desired accuracy.
For a given numerical scheme, the numerical error $e$ is proportional to $(DOF)^P$, where $P$ represents the order of the method \cite{wang2016perspective}. To achieve comparable errors between high-order (HO) and low-order (LO) methods (i.e. $e_{HO} \approx e_{LO}$), the high-order mesh can be coarsened according to the relation: $(DOF)_{HO} = \exp\left(\frac{P_{LO}}{P_{HO}} \cdot \log[(DOF)_{LO}]\right)$. For example, selecting $P_{LO} = 2$ and $P_{HO} = 5$, a low-order mesh with $(DOF)_{LO} = 100$ million corresponds to a high-order mesh with $(DOF)_{HO} = 1.6$ thousand, showing a reduction by a factor of 60,000 for the same precision. 
Although this estimate is optimistic, since small cells are still necessary near walls and geometric features, it nonetheless highlights the potential of high-order methods to drastically reduce the number of degrees of freedom without sacrificing accuracy. Moreover, additional accuracy can be obtained with minimal computational cost by employing local p-refinement, where only specific mesh elements utilize high-order polynomials \cite{rubio2015quasi,kompenhans2016adaptation,kompenhans2016comparisons,rueda2019truncation,rueda2019p,ntoukas2021free}.



To our knowledge, this is the first study to compare the h- and p-refinement methods in the context of wind turbine wake development. 
This study will provide a comprehensive comparison of these two methods, evaluating their effectiveness, computational efficiency, and suitability for various scenarios in the field of wind energy. The findings of this research could potentially guide future studies and applications in wind turbine wake modeling, thus contributing to the optimization of the wind farm using high-order h/p methods. 

The remainder of the paper is organized as follows. Section \ref{sec:meth} summarizes the high-order discontinuous Galerkin method used, including local p- and h-refinement through mortar interfaces. Then, Section \ref{sec:wind_turbine} introduces the wind turbine case and the results obtained for a variety of meshes. Conclusions follow in Section \ref{sec:conc} .

\section{Methodology}\label{sec:meth}

\subsection{HORSES3D a high-order discontinuous Galerkin h/p solver}
All developments described in this work are implemented in the open-source solver Horses3D  \cite{horses3d_paper}, developed at the school of aeronautics ETSIAE-UPM in Madrid, and available on Github (\url{https://github.com/loganoz/horses3d}). 
This solver is a high-order discontinuous Galerkin tool capable of addressing a wide array of flow applications. These applications encompass compressible flows (with or without shock waves), incompressible flows, a range of RANS and LES turbulence models, particle dynamics, multiphase flows, and aeroacoustics. Furthermore, Horses3D is equipped to manage body-fitted, immersed boundaries, and actuator lines.

%
One of the main advantages of DG methods is their ability to accurately capture high-order spatial and temporal variations of the solution, which makes them particularly suitable for simulating flows with sharp gradients and complex flow phenomena. DG methods also exhibit good numerical stability and conservation properties because of the local nature of the approximation and the use of fluxes at the interfaces of the elements. 

 In this section, we provide only a brief overview of the fundamental concepts of DG discretizations for the compressible Navier-Stokes retained in this work; see \ref{sec:cNS}.\\

The physical domain is discretized into non-overlapping curvilinear hexahedral elements, denoted by $e$, which are geometrically mapped to a reference element, $el$, through a polynomial transfinite transformation. This mapping establishes the relationship between the physical coordinates $\vec{x}$, and the local reference coordinates $\vec{\xi}$. The system of equations is then transformed using this mapping, yielding the following expression:
\begin{equation}
\boldsymbol{q}_t  + \nabla_x\cdot\ssvec{F}_e = \nabla_x\cdot\ssvec{F}_{v,turb}+\boldsymbol{S({q}_t)},
\rightarrow
J \boldsymbol{q}_t  + \nabla_\xi\cdot\cssvec{F}_e = \nabla_\xi\cdot\cssvec{F}_{v,turb}+J \boldsymbol{S({q}_t)},
\label{eq:compressibleNScompact_transformed}
\end{equation}
where $J$ is the Jacobian of the transfinite mapping, $\nabla_\xi$ is the differential operator in the reference space and $\cssvec{F}$ are the contravariant fluxes \cite{2009:Kopriva}. The source term $\mathbf{S}$ will be used to incorporate the actuator line forcing for the rotating blades and immersed boundaries for the tower and nacelle \cite{marino2024modelling}.
To derive DG schemes, we multiply Eq.~\eqref{eq:compressibleNScompact_transformed} by a locally smooth test function $\phi_j$ ($0\leq j\leq P$, where $P$ is the polynomial degree) and integrate over an element $el$ to obtain the weak form:
\begin{equation}\label{eq::NS2}
\int_{el}J \boldsymbol{q}_t\phi_j+\int_{el} \nabla_\xi\cdot\cssvec{F}_e\phi_j  =\int_{el} \nabla_\xi\cdot\cssvec{F}_{v,turb}\phi_j+\int_{el}J\boldsymbol{S({q}_t)}\phi_j.
\end{equation}

By integrating the term with the inviscid fluxes, $\mathbf{F}_e$, by parts, we obtain a local weak form of the equations (one per mesh element) with the boundary fluxes separated from the interior:
\begin{equation}\label{eq::NS3}
\int_{el}J \boldsymbol{q}_t\phi_j +  \int_{\partial el} \cssvec{F}_e\cdot{\mathbf{n}}\phi_j-\int_{el} \cssvec{F}_e\cdot\nabla_\xi\phi_j
=\int_{el} \nabla_\xi\cdot\cssvec{F}_{v,turb}\phi_j+\int_{el}J\boldsymbol{S({q}_t)}\phi_j,
\end{equation}
where ${\mathbf{n}}$ is the unit outward vector of each face of the reference element ${\partial el}$.
Discontinuous fluxes at inter-element faces are replaced by a numerical inviscid flux $\mathbf{F}_{e}^{\star}$, to couple the elements:
\begin{equation}\label{eq::NS5}
\int_{el}J \boldsymbol{q}_t\phi_j + \underbrace{{ \int_{\partial el} \cssvec{F}_{e}^{\star}}\cdot{\mathbf{n} \phi_j}-\int_{el} \cssvec{F}_e\cdot\nabla_\xi\phi_j}_\text{Convective fluxes}
= \underbrace {\int_{el} (\nabla_\xi\cdot\cssvec{F}_{v,turb})\phi_j }_\text{Viscous and Turbulent fluxes} +\underbrace{ \int_{el}J\boldsymbol{S({q}_t)}\phi_j}_\text{Source term AL-IBM}.
\end{equation}

The equations for each element are coupled through Riemann fluxes $\mathbf{F}_{e}^{\star}$. Viscous terms are also integrated by parts but require further manipulations to obtain usable discretisations \cite{unified,horses3d_paper}. Nonlinear inviscid and viscous numerical fluxes (including turbulent ones) can be chosen appropriately to control dissipation in the numerical scheme~\cite{Manzanero_2020,Ferrer_2017,jumpKou}. In this work, we use the Lax-Friederich flux and Bassi-Rebay 1 (BR1) schemes, respectively. 

In a final step, we approximate the numerical solution and fluxes using polynomials (of order $P$) and evaluate all integrals using Gaussian quadrature rules. Specifically, we use Gauss-Legendre points for the quadrature in this work. In addition, we have complemented the compressible Navier-Stokes equations with the Vreman \cite{Vreman_2004} and Smagorinsky \cite{Smagorinsky_1963} Large Eddy Simulation subgrid models.  To advance the solution in time, we use an explicit Runge-Kutta 3 time marching scheme (with a $\text{CFL}<1$ in all cases). Finally, it is important to note that when selecting a polynomial of order $P$, the formal order of the scheme is $P+1$ (i.e., how rapid the error decay is when refining the mesh). In the next sections, we perform simulations using $P=1$ (second order scheme) to $6$ (seventh order accuracy).

\subsection{Treatment of non-conforming interfaces}

The methodology described in the previous section is adequate for addressing problems that utilize a uniform polynomial order across the entire domain and when each element is connected to only one neighboring element. However, in some cases, it may be desirable to achieve high accuracy in specific regions of the flow (e.g., around the turbine and in the wake) while allowing for lower accuracy in other regions (e.g., the surrounding external domain). In such scenarios, it becomes necessary to define mortar connections at the interface between elements.


In Horses3D, the connection between non-conforming elements is achieved using the mortar method, as described in \cite{ref117}, \cite{ref118}. The fundamental concept behind the mortar method is analogous to the use of mortar (the ``cement'') to connect neighboring elements (the ``bricks''). In this approach, the 3-dimensional elements communicate indirectly via an intermediate 2-dimensional structure, known as the mortar, rather than directly communicating with neighboring elements. Mortars allow the recovery of high-order accuracy when gluing together elements that have different polynomial orders (local p-refinement) and also when hanging nodes are present (local h-refinement).

While the connection of polynomial orders is generally flexible, the non-conforming connections in 3D simulations are limited to a 4-to-1 ratio. Further details on the theoretical background and implementation are provided in \ref{sec:mortar}. 

\section{Wind turbine case: comparison of h- and p- refinement}\label{sec:wind_turbine}
\subsection{Wind turbine case}

\begin{figure}[t]
    \centering
           \hspace*{-5.0cm}
    \begin{subfigure}[b]{1.8\textwidth}  
        \centering
        \includegraphics[width=\textwidth]{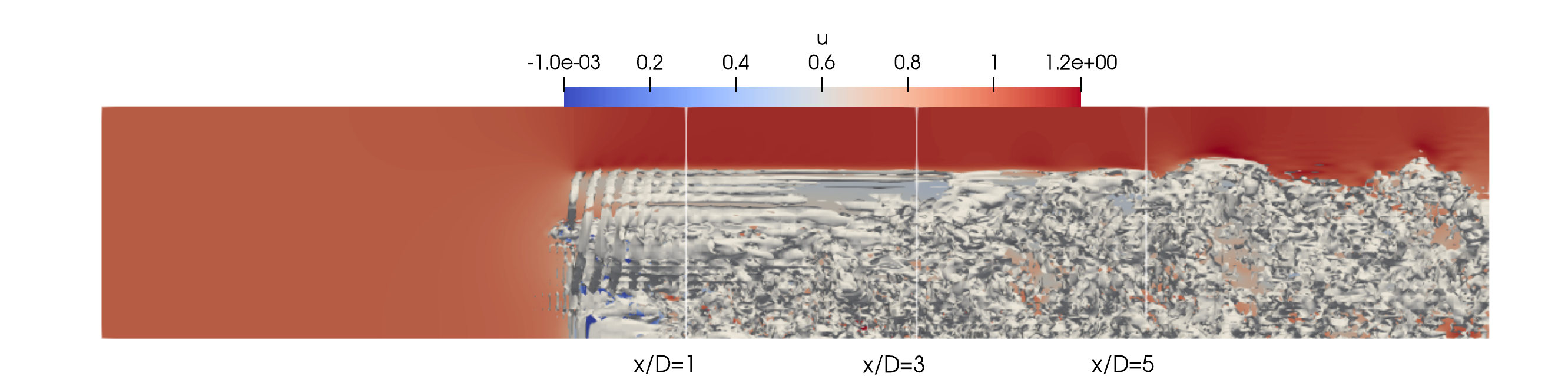}
        \caption{Streamwise velocity contours on a plane perpendicular to the rotor and three dimensional Q-criterion isocontours.}    \label{fig:ref}
    \end{subfigure}
    
    \vskip\baselineskip  

    \begin{subfigure}[b]{1.0\textwidth}  
        \centering
        \includegraphics[width=\textwidth]{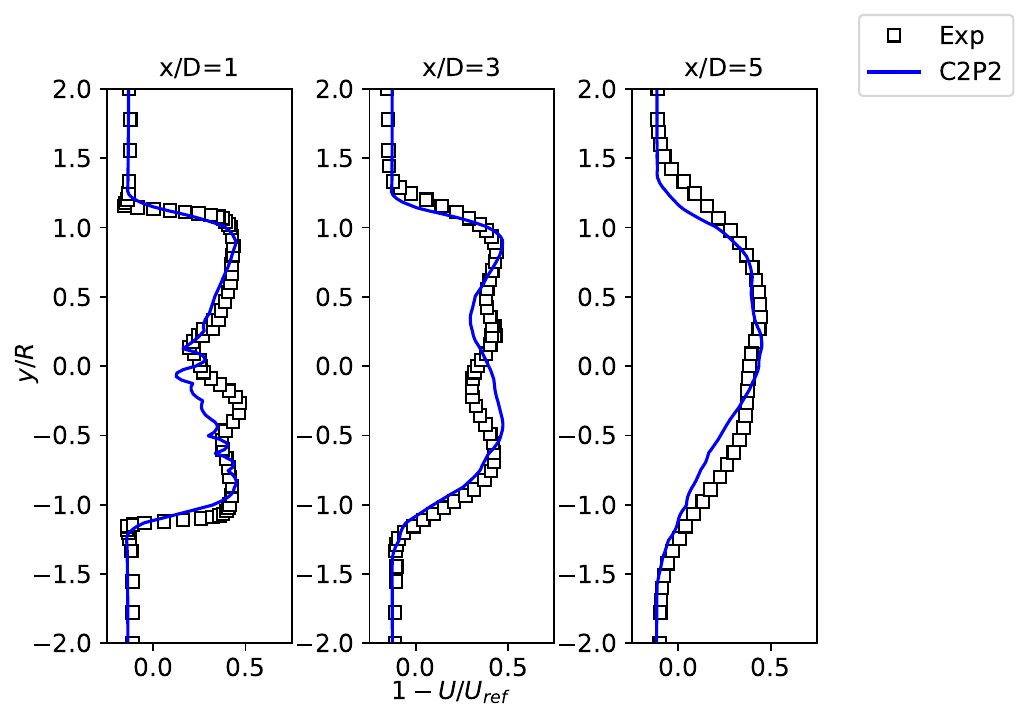}
        \caption{Horizontal profiles of mean streamwise velocity deficit at three down-stream position $x/D=1$, 3 and 5 and comparison with experimental data \cite{krogstad2013blind}.} \label{fig:C2P2}
    \end{subfigure}
    
    \caption{Fine simulation using mesh C2P2. See next section and Table \ref{tab:sum_setup} for details on the mesh.}
   
\end{figure}

We now compare h- and p- refinement for the three-dimensional turbulent flow past a wind turbine at $Re_c=103600$ (based on the blade chord). This flow has been experimentally tested at the Norwegian University of Science and Technology, the wind turbine has a diameter $D=0.894$m, the blades are made up of NREL S826 airfoils \cite{krogstad2013blind}. The blind test used a wind tunnel of dimensions $[L \times W \times H] = [11.15 \times 2.71 \times 1.8]$ m, a low turbulent intensity of $0.3 \%$, and a uniform inflow velocity. Various tip speed ratios ($\delta$) were used in the blind test. In this study, we used the optimal tip speed ratio $\delta=\gamma D/2U_{\infty} = 6$, with $U_{\infty}=10$~m/s and $\gamma=134.228$~rad/s. The blade tip Reynolds number for this case is $Re_c = \delta U_{\infty} c_{tip}/\nu= 103600$, where $c_{tip}=0.025926$ m is the tip chord length and $\nu$ is the kinematic viscosity of air. An immersed boundary method \cite{KOU2022110721,KOU2022110817,horses3d_paper} has been used to model the tower and the nacelle, the rotating blades were modeled using an actuator line method \cite{sorensen2002numerical,marino2024modelling,BOTEROBOLIVAR2024120476}. The Mach number is set to $M=0.03$, with free-slip boundary conditions applied at the wind tunnel walls, and no inlet turbulence is introduced. Unless otherwise specified, Large eddy simulations are performed using the Vreman sub-grid scale closure model, see \ref{sec:cNS} for details.
The simulations are conducted for $T=2\,s$. After that, the statistics are then gathered within an interval of length of $0.5\,s$. 
Figure \ref{fig:ref} shows the stream velocity magnitude, with white lines indicating the locations at \(x/D = 1\), \(3\), and \(5\). Figure \ref{fig:C2P2} illustrates the horizontal profiles of the time-averaged streamwise velocity deficit at various downstream positions.




\clearpage
\subsection{Mesh (hp) descriptions and summary of results}

\begin{figure}[h!]
    \centering
    \begin{subfigure}[b]{1.7\textwidth}  
        \centering
        \hspace*{-9.0cm}
        \includegraphics[width=\textwidth]{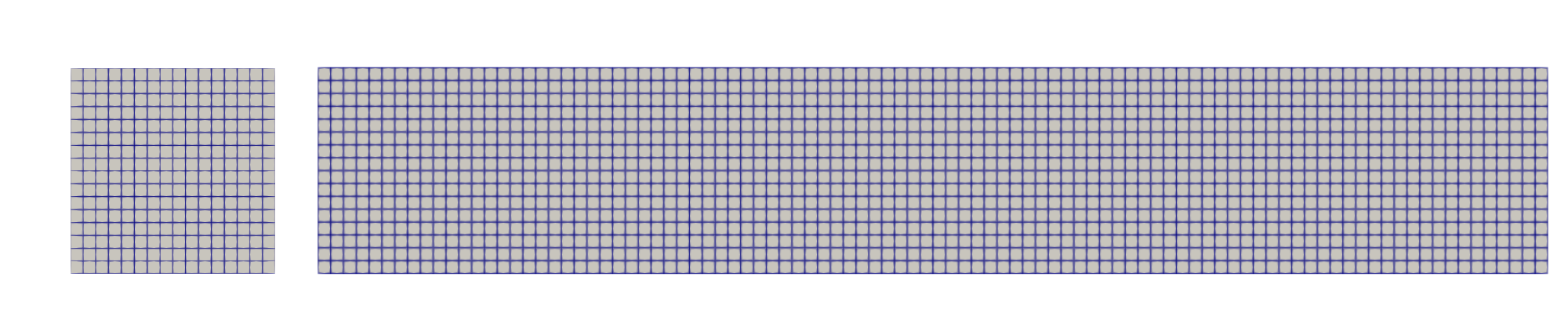}
        \caption{\hspace*{9.0cm}}
    \end{subfigure}
    
    \vskip\baselineskip  

    \begin{subfigure}[b]{1.7\textwidth}  
        \centering
        \hspace*{-9.0cm}
        \includegraphics[width=\textwidth]{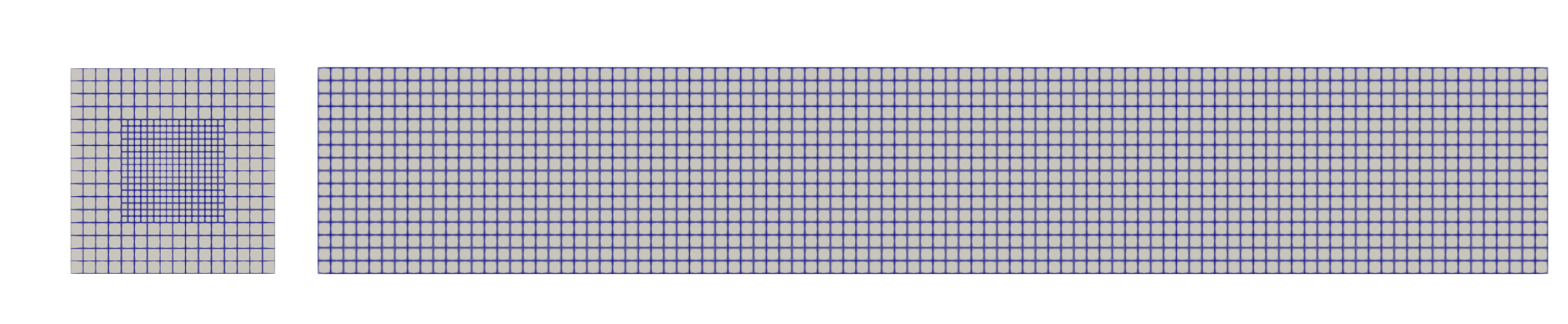}
        \caption{\hspace*{9.0cm}}
    \end{subfigure}

    \vskip\baselineskip  

    \begin{subfigure}[b]{1.7\textwidth}  
        \centering
        \hspace*{-9.0cm}
        \includegraphics[width=\textwidth]{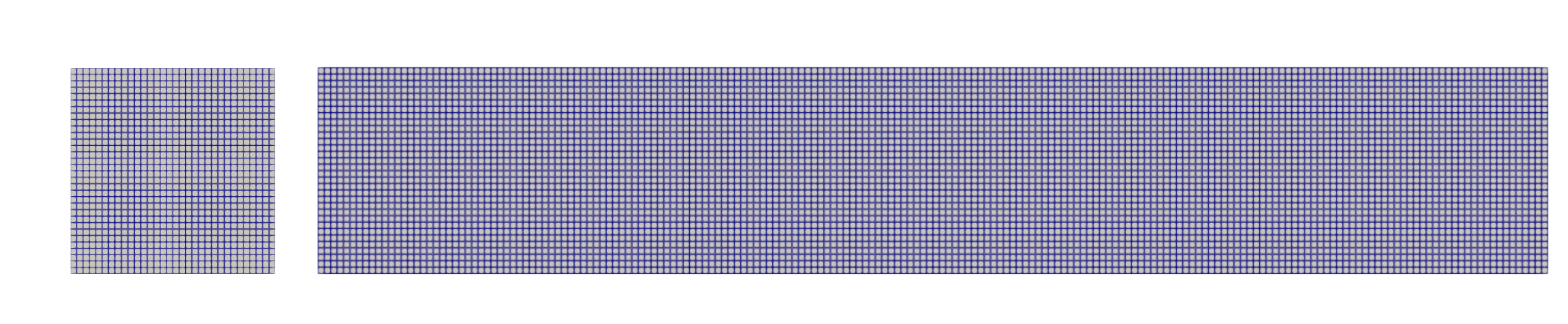}
        \caption{\hspace*{9.0cm}}
    \end{subfigure}
    
    \caption{Visualization of the three meshes used in this study. (a) y/z-plane (left) and x/y-plane (right) for C1. (b) y/z-plane (left) and x/y-plane (right) for NC. (c) y/z-plane (left) and x/y-plane (right) for C2.}
    \label{fig:meshes}
\end{figure}

In this study, two different conforming h-meshes with Cartesian grids are considered: one coarse mesh (C1) and one fine mesh (C2). The coarse mesh (C1) consists of $[16 \times 16 \times 96]$ elements, while the fine mesh (C2) consists of $[32 \times 32 \times 192]$ elements. The $D/\Delta x$ ratio, where $D$ is the turbine diameter, is approximately $15.9$
for the fine mesh and $7.9$ 
for the coarse one. It is important to note that, since a high-order method is employed, spatial resolution is further enhanced by increasing the polynomial order.
Additionally, an intermediate non-conforming mesh (NC) is considered, where refinement is applied only in the central region (where the blades and wake are located). The central region consists of $[16 \times 16 \times 192]$ elements that correspond to the element size of the fine mesh, while the outer region consists of 18432 elements that correspond to the coarse mesh size. This configuration reduces the element size only in the regions of interest, leading to a reduction in computational cost.
The three meshes are depicted in Figure \ref{fig:meshes}, where the variations in mesh size can be seen for each region.

In this study, polynomial orders ranging from $P=1$ to $6$ (second to seventh order) are considered to increase the resolution within each element. Various strategies are employed to distribute the different polynomial orders. The configurations used are summarized in Table \ref{tab:sum_setup}. The notation "Mesh--pol order" is adopted to describe each setup. In some cases, two polynomial orders are applied within the same problem: a higher polynomial order in the wake-blades region and a lower one in the outer region. The region with higher polynomial order corresponds to the refined area of the NC mesh described in the previous paragraph. For these cases, the notation "Mesh--high pol order in wake-blades region--low pol order outside" is used. 

\begin{table}[h!]
\centering
\begin{tabular}{|c|c|}
\hline
\textbf{Mesh} & \textbf{nDOFs} \\
\hline
C1P1      & 196,608    \\
C1P2      & 663,552    \\
C1P3      & 1,572,864  \\
C2P1      & 1,572,864  \\
C2P2      & 5,308,416  \\
C2P3P1    & 4,325,376  \\
NCP1      & 540,672    \\
NCP2      & 1,824,768  \\
C1P2P1    & 313,344    \\
C1P3P1    & 540,672    \\
C1P4P1    & 915,456    \\
C1P5P1    & 1,474,560  \\
C1P6P2    & 2,605,056  \\
\hline
\end{tabular}
\caption{Summary of Mesh Configurations ("Mesh–pol order") and number of degrees of freedom (nDOFs)}
\label{tab:sum_setup}
\end{table}

\begin{figure}[h!]
       \begin{center}
       \hspace*{-1.5cm}
\includegraphics[scale=0.54]{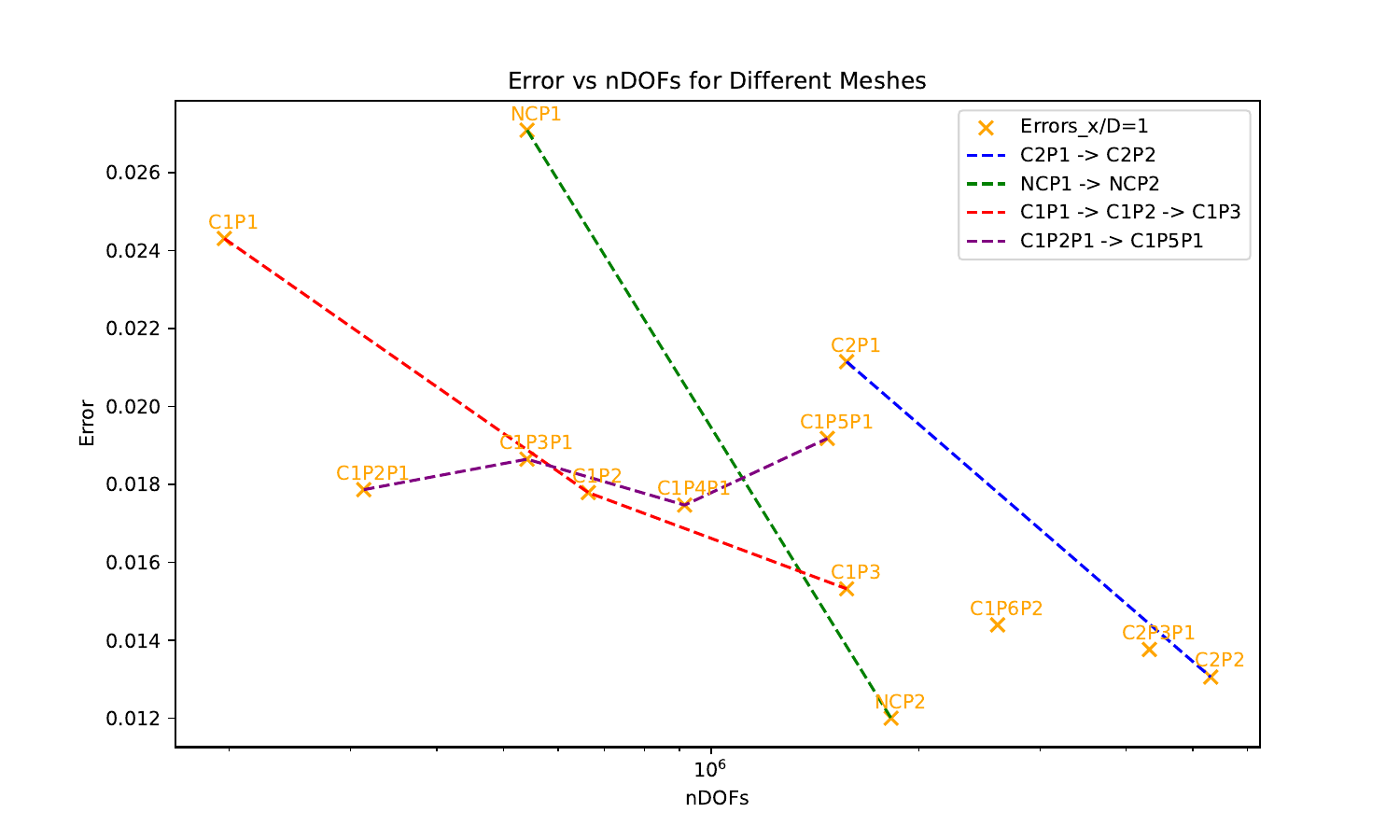}
\end{center}
\caption{Summary of errors vs nDOFs for all meshes, at $x/D = 1$. }
\label{fig:error1}
\end{figure}

\begin{figure}[h!]
       \begin{center}
       \hspace*{-1.5cm}
\includegraphics[scale=0.54]{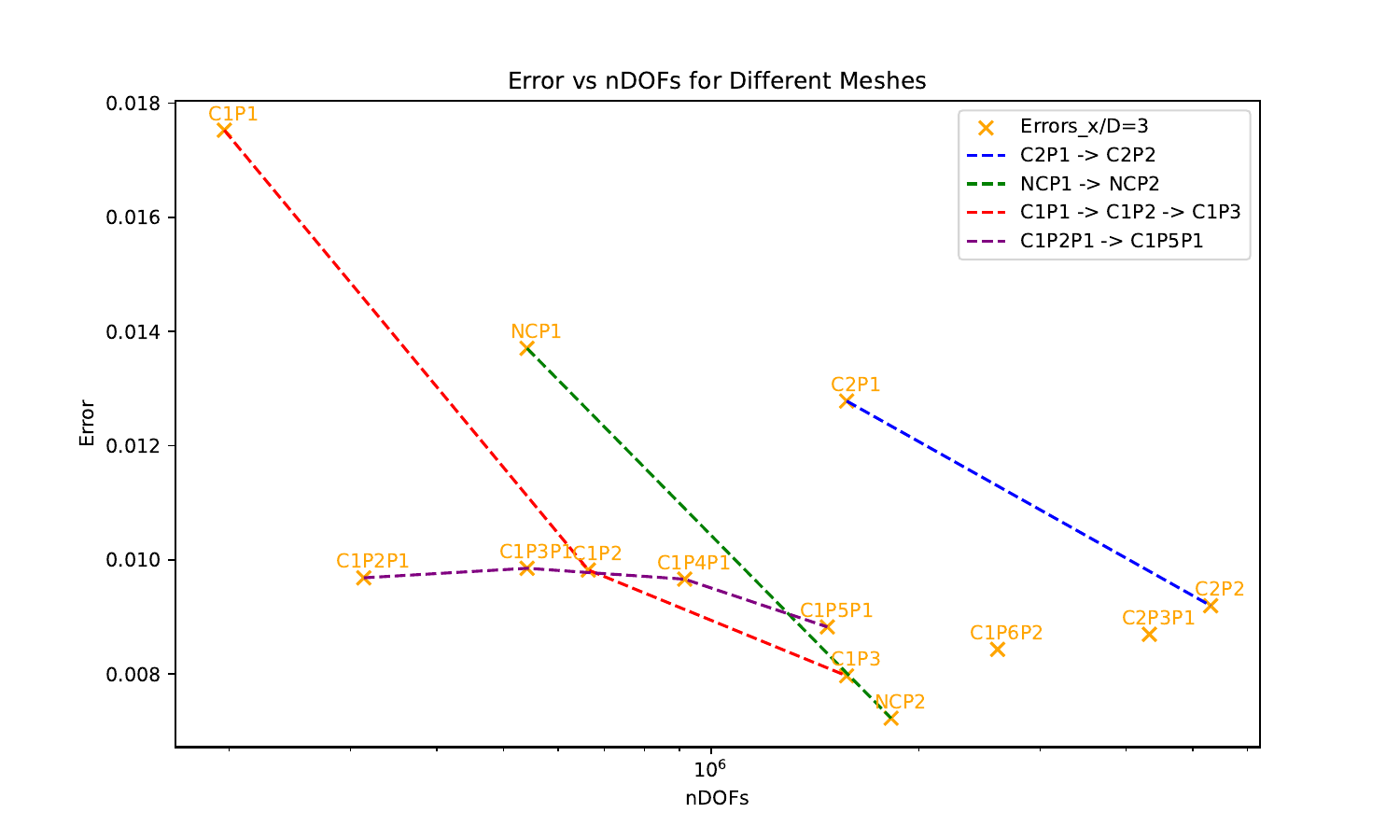}
\end{center}
\caption{Summary of errors vs nDOFs for all meshes, at $x/D = 3$.  }
\label{fig:error3}
\end{figure}

\begin{figure}[h!]
       \begin{center}
       \hspace*{-1.5cm}
\includegraphics[scale=0.54]{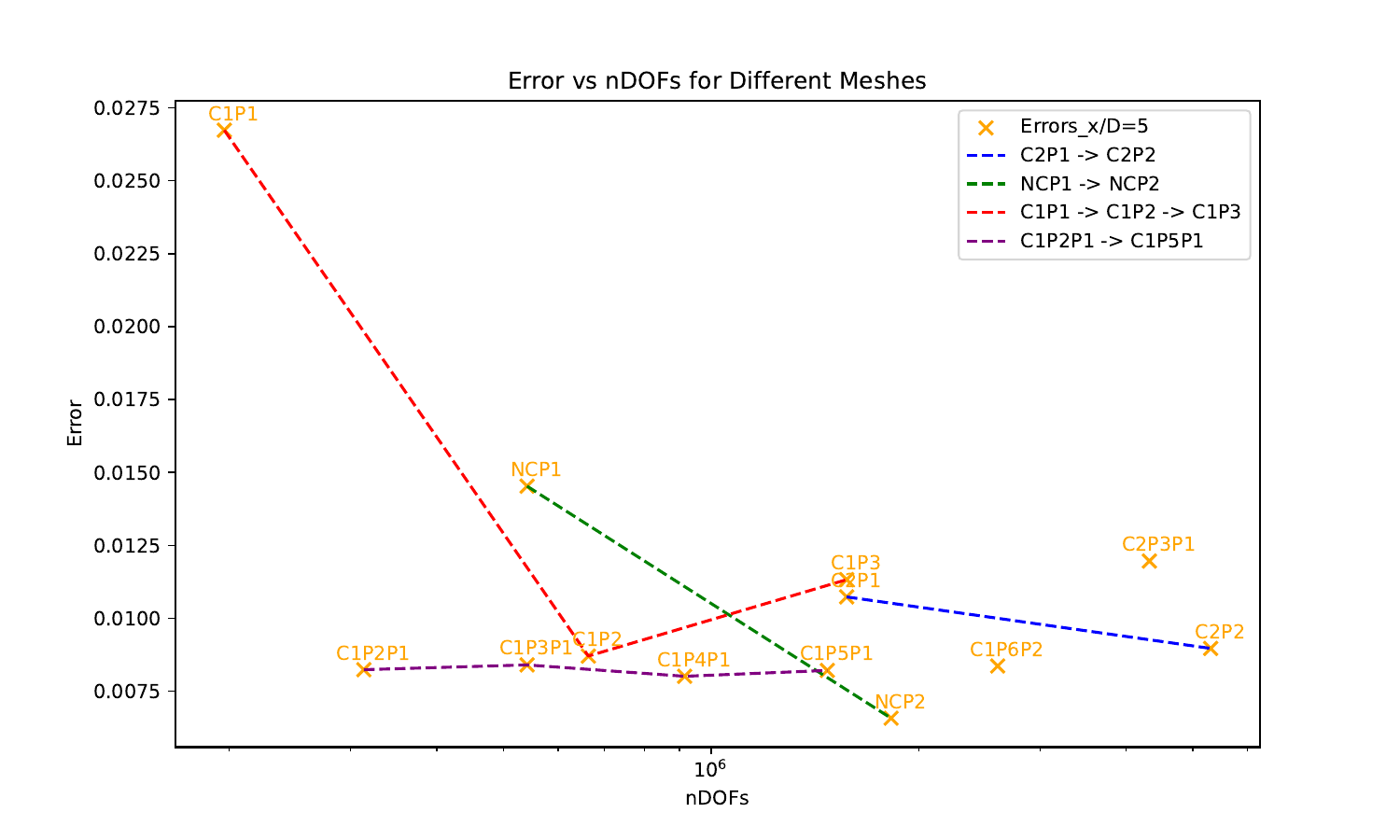}
\end{center}
\caption{Summary of errors vs nDOFs for all meshes, at $x/D = 5$. }
\label{fig:error5}
\end{figure}

Figures \ref{fig:error1}, \ref{fig:error3}, and \ref{fig:error5} provide a summary of the results obtained in this study. The number of degrees of freedom (nDOFs) is plotted against the error relative to the experimental results of \cite{krogstad2013blind} for various $x/D$ positions. The error is quantified using the $L_2$ size-normalized norm between the simulation results and the experimental data, defined as:
\begin{equation}
    \text{Error} = \frac{1}{N}\sum_{i=1}^N (u_{exp}[i] - u_{num}[i])^2,
\end{equation}
where $N$ represents the number of experimental points for each $x/D$ position, $u_{exp}$ is the experimentally measured average velocity deficit, and $u_{num}$ is the averaged numerical result at the same spatial location. 
In general, it is evident that increasing the DOFs results in a numerical solution that is more closely aligned with the experimental data. However, the error reduction tends to plateau beyond a certain resolution, likely due to modeling limitations, such as the use of an actuator line model for the blades or the tower being represented by an immersed boundary method. For a more detailed discussion of the modeling errors in our approach, refer to \ref{sec:modelingerrors}.  We also observe that further from the rotor (e.g., at $x/D=5$) the effect of mesh refinement is less noticeable, since there are smaller variations of the errors when refining. This observation suggests that the meshes selected for the study are fine enough to capture the correct physics of wake development far from the rotor plane. 

\begin{figure}[h!]
       \begin{center}
       \hspace*{-1.5cm}
\includegraphics[scale=0.54]{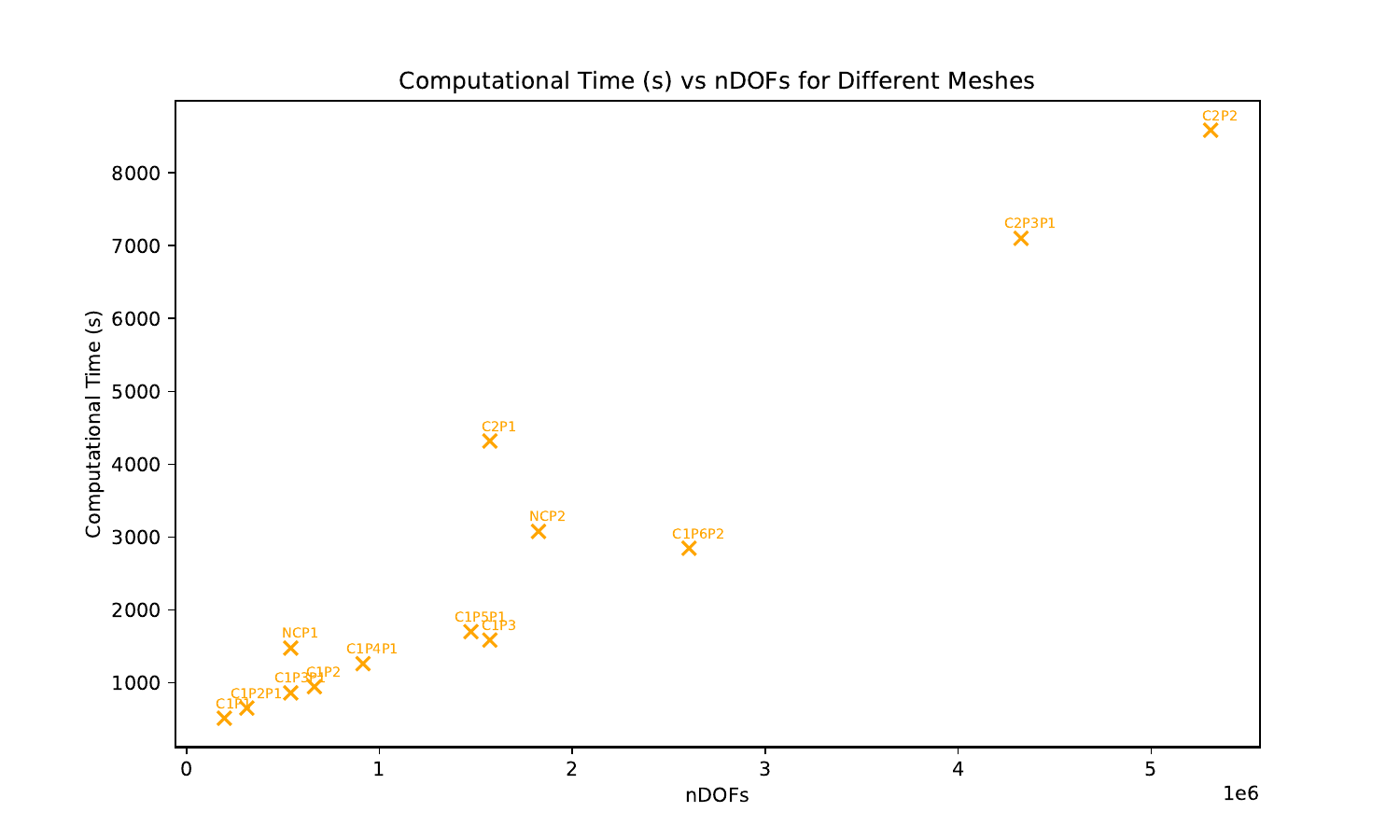}
\end{center}
\caption{Summary of computational cost (serial simulation run for 100 iterations) vs nDOFs for all meshes. }
\label{fig:cost}
\end{figure}

Finally, Figure \ref{fig:cost} summarizes the computational cost relative to the number of degrees of freedom for each mesh. As anticipated by theory, the cost increases linearly with the number of degrees of freedom. Notably, due to implementation and hardware specifics, the p refinement demonstrates slightly better scaling compared to the h refinement. Additionally, the error reduction is faster with p-refinement than with h-refinement. Overall, we conclude that p-refinement is more efficient than h-refinement. In the following sections, we will provide a detailed analysis of the results, emphasizing the comparison between these two refinement strategies.

\clearpage

\subsection{Assessment of p-refinement vs h-refinement}

\begin{figure}[h!]
       \begin{center}
\includegraphics[scale=0.6]{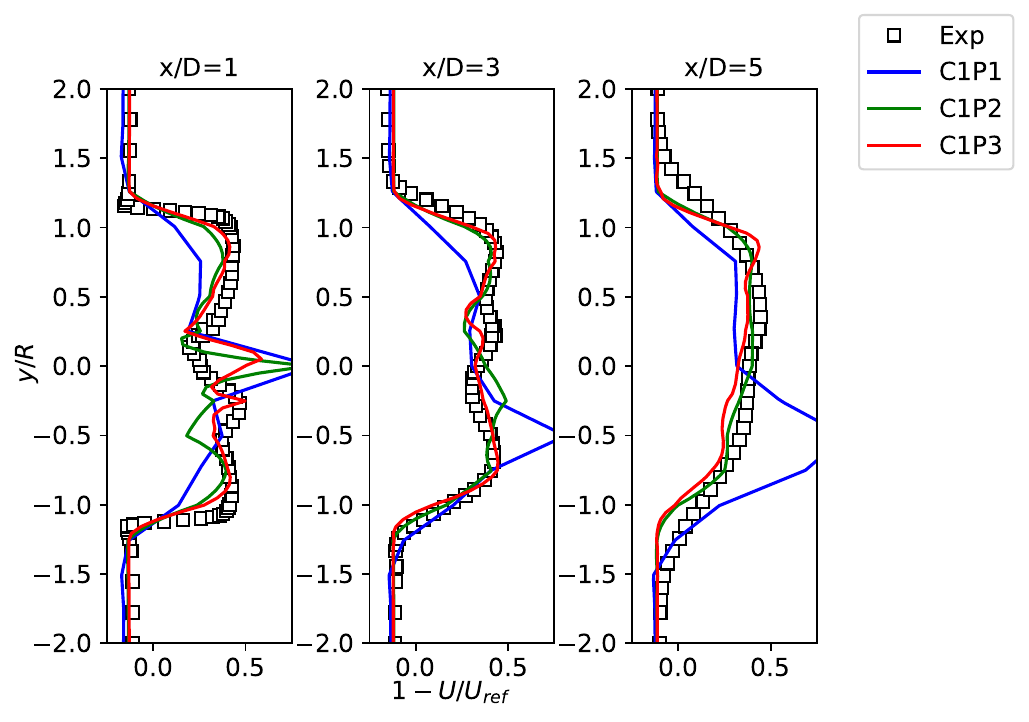}
\end{center}
\caption{Horizontal profiles of mean streamwise velocity deficit at three down-stream positions \begin{math} x/D \end{math} for \begin{math} P \in \{1,2,3\} \end{math} with coarse mesh C1 and comparison with experimental data. }
\label{fig:C1_P1vsP2vsP3}
\end{figure}
\begin{figure}[h!]
       \begin{center}
\includegraphics[scale=0.6]{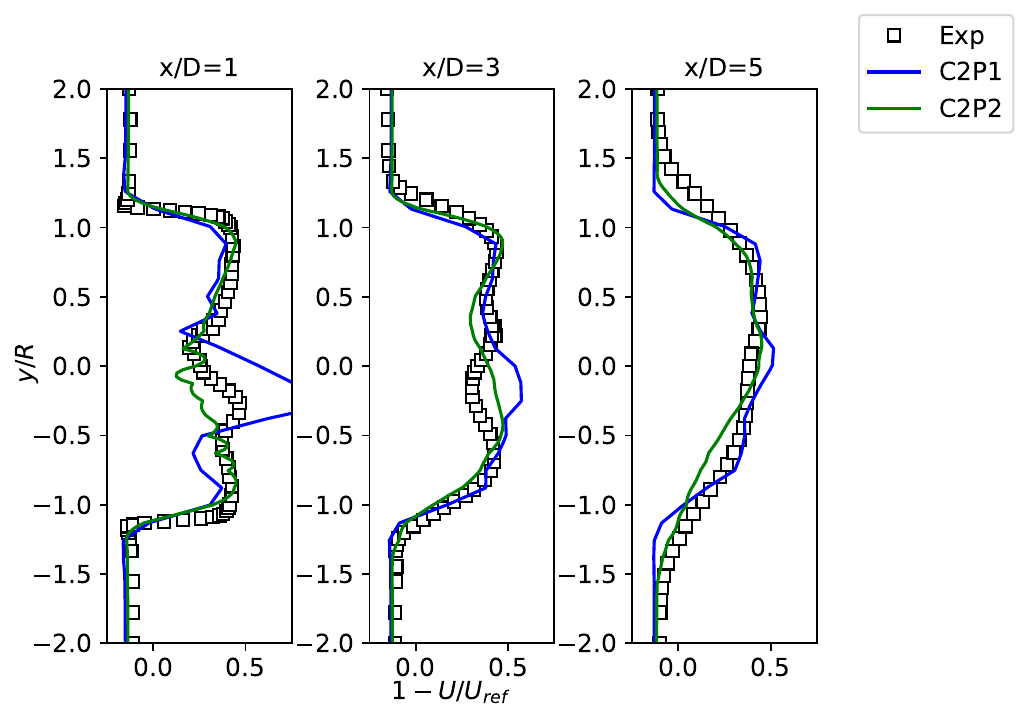}
\end{center}
\caption{Horizontal profiles of mean streamwise velocity deficit at three down-stream positions \begin{math} x/D \end{math} for \begin{math} P \in \{1,2\} \end{math} with fine mesh C2 and comparison with experimental data.}
\label{fig:C2_P1vsP2}
\end{figure}

We first focus on the conforming test cases and compare the performance of the p-refinement against the h-refinement. 
In Figures \ref{fig:C1_P1vsP2vsP3} and \ref{fig:C2_P1vsP2}, we observe the effect of increasing the polynomial order on both coarse and fine meshes (C1 and C2). As anticipated, the results demonstrate significant improvement when the polynomial order is increased, underscoring the enhanced accuracy attainable through higher-order polynomials. This improvement is attributed to the greater flexibility of higher-order polynomials in approximating complex solution features, such as large gradients and flow structures within the computational domain. In the coarse mesh (C1), the enhancement is particularly noticeable, indicating that higher polynomial orders can compensate for the coarse mesh. In contrast, in the fine mesh (C2), the results converge more rapidly, suggesting that the combination of a finer mesh and higher polynomial order yields enhanced accuracy. By enriching the solution space, higher polynomial orders enable a more precise representation of the underlying physical phenomena. 
However, this increase in accuracy is accompanied by a higher computational cost. 
The results presented here highlight the efficacy of p-adaptation in capturing detailed flow characteristics, which may be inadequately resolved by lower-order methods.

\begin{figure}[h!]
       \begin{center}
\includegraphics[scale=0.6]{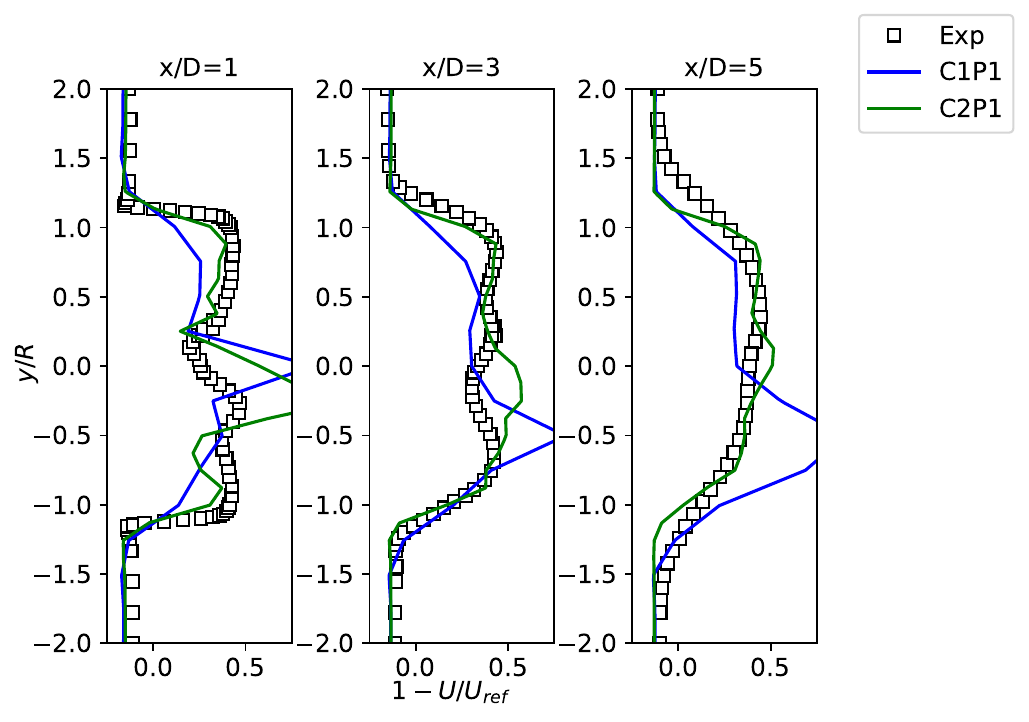}
\end{center}
\caption{Horizontal profiles of mean streamwise velocity deficit at three down-stream positions \begin{math} x/D \end{math} for \begin{math} P=1 \end{math} with both fine and coarse meshes C1 and C2 and comparison with experimental data.}
\label{fig:P1_C1vsC2}
\end{figure}

\begin{figure}[h!]
       \begin{center}
\includegraphics[scale=0.6]{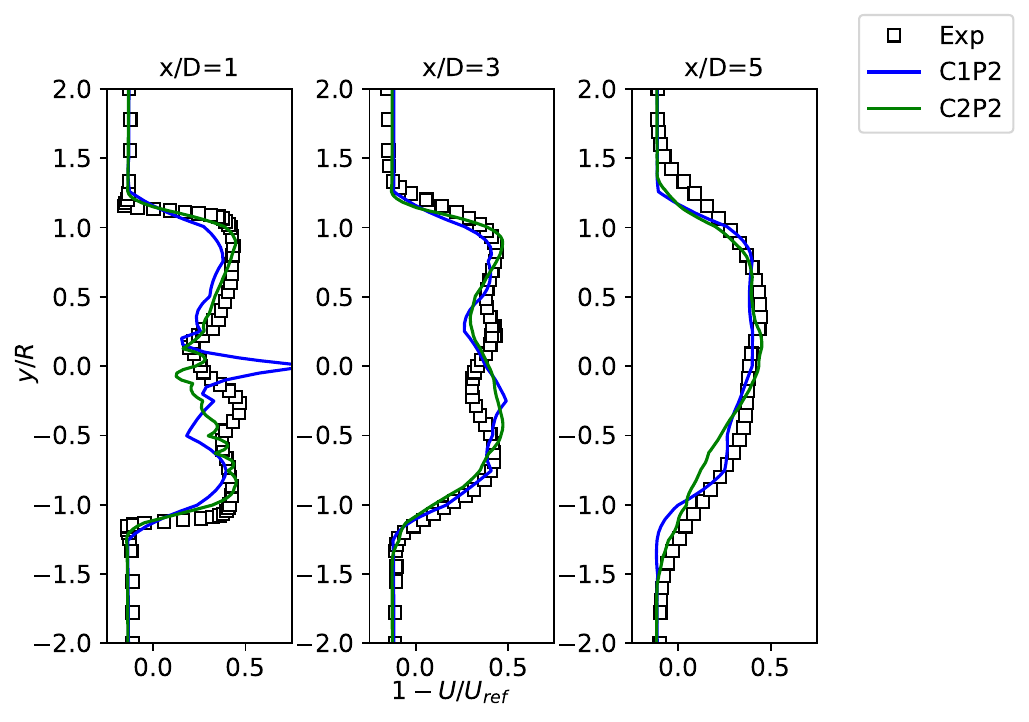}
\end{center}
\caption{Horizontal profiles of mean streamwise velocity deficit at three down-stream positions \begin{math} x/D \end{math} for \begin{math} P=2 \end{math} with both fine and coarse meshes C1 and C2 and comparison with experimental data.}
\label{fig:P2_C1vsC2}
\end{figure}

In Figures \ref{fig:P1_C1vsC2} and \ref{fig:P2_C1vsC2}, we observe the effect of increasing the number of elements (h-refinement) with polynomial orders 1 and 2. As expected, increasing the number of elements leads to a significant improvement in accuracy. This enhancement is due to the finer mesh’s ability to resolve smaller-scale features and gradients within the computational domain more effectively. With polynomial order $P=1$, the improvement is evident as the mesh becomes finer, capturing more details of the flow structure that are otherwise missed in coarser grids. When the polynomial order is increased to $P=2$, the results further improve, particularly in the finer meshes, indicating that the h-refinement in conjunction with higher polynomial orders yields superior accuracy.

However, this increase in accuracy comes with a corresponding increase in computational cost as a result of the larger number of elements and the increased degrees of freedom. The results presented here highlight the importance of balancing mesh density and computational efficiency. They also demonstrate the efficacy of h-refinement in enhancing solution accuracy, particularly when combined with moderate polynomial orders, making it a crucial strategy for capturing intricate physical phenomena that lower-resolution meshes and lower-order polynomials might fail to resolve.

To assess the effectiveness of h-refinement versus p-refinement in reducing error, we refer to Figures \ref{fig:error1}, \ref{fig:error3}, and \ref{fig:error5}. It is clear that the rate of convergence is higher when the polynomial order is increased than when the mesh is refined. Furthermore, for a similar number of degrees of freedom, utilizing a higher polynomial order with a coarser mesh is more efficient than employing a lower order with a finer mesh. This is evident in the comparison between C2P1 and C1P3, particularly near the wind turbine at \(x/D = 1\) and \(x/D = 3\). However, at \(x/D = 5\), the error stabilizes around $1\%$, with only minimal improvements observed in all cases.

\begin{figure}[h!]
    \begin{center}
        \hspace*{-3.5cm}
        \includegraphics[scale=0.45]{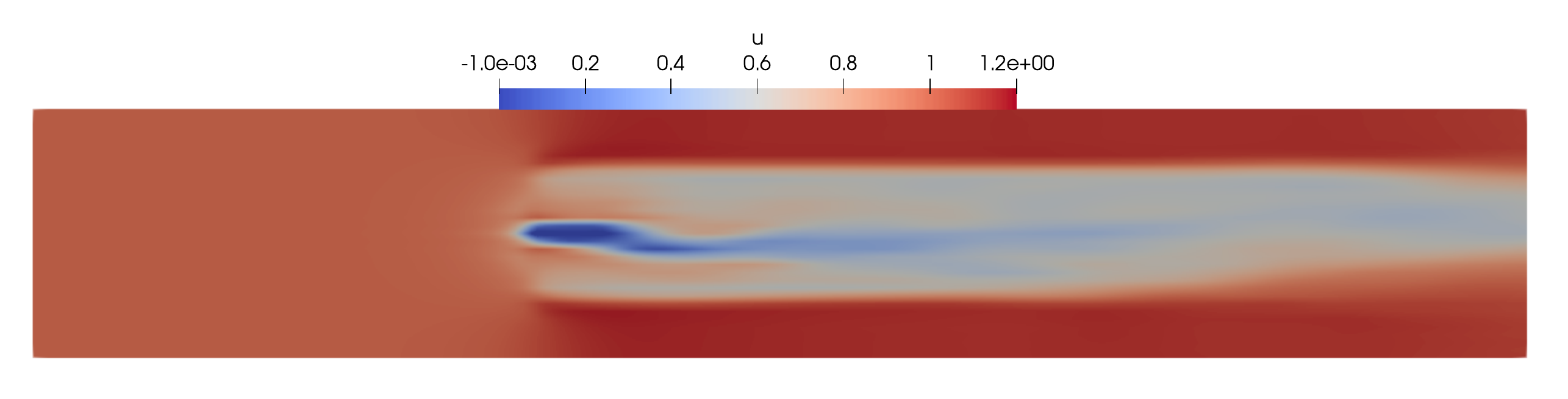}
        
        \vspace{0.5cm} 
        
        \hspace*{-3.5cm}
        \includegraphics[scale=0.45]{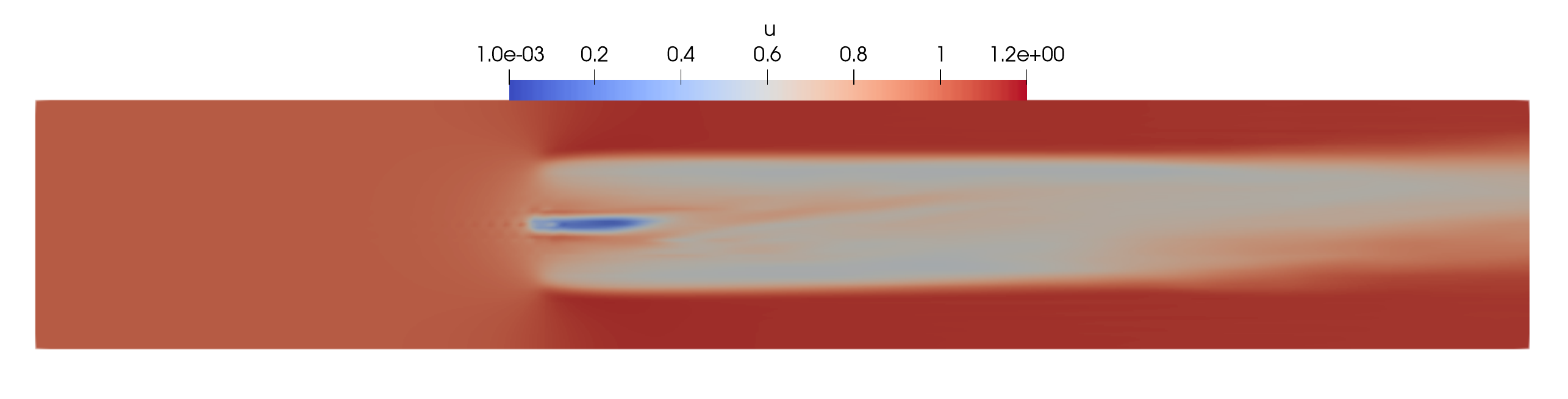}
        
        \vspace{0.5cm} 
        
        \hspace*{-3.5cm}
        \includegraphics[scale=0.62]{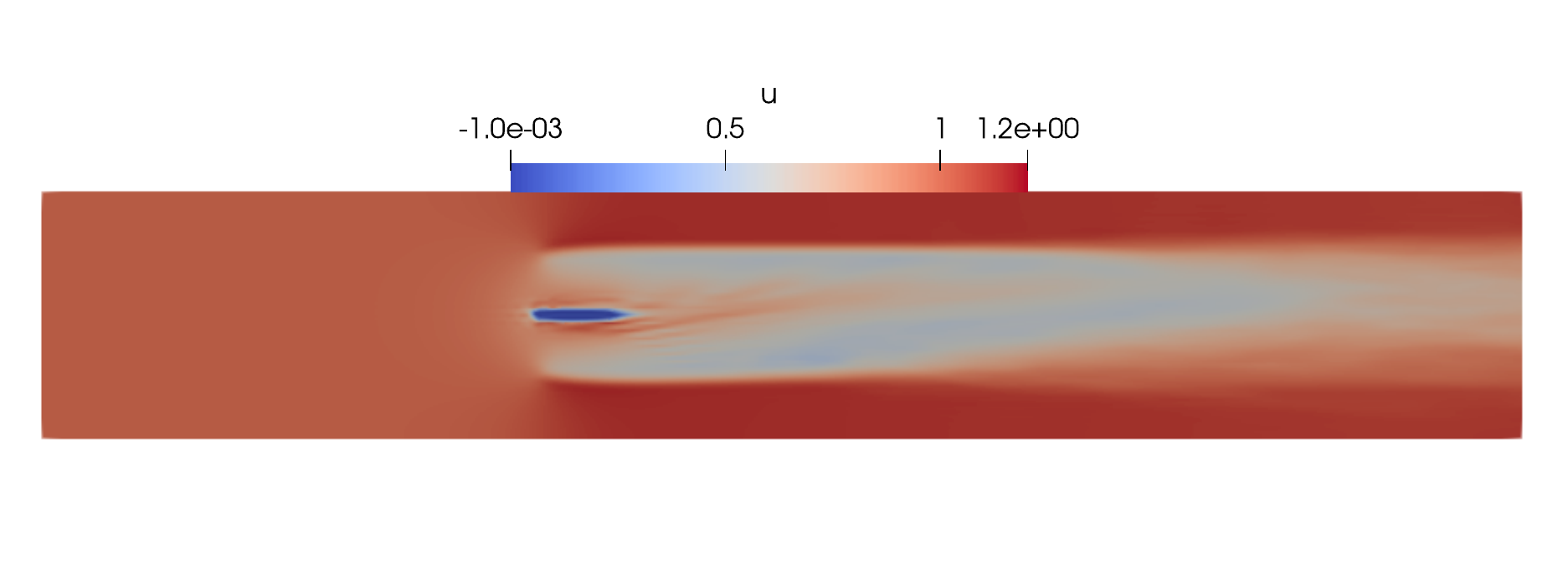}
    \end{center}
    \caption{Contours of streamwise velocity for different meshes: (TOP) C2P1, (MIDDLE) C1P3, and (BOTTOM) C2P2 (reference). Each plane is through the wind turbine rotor, viewed from above.}
    \label{fig:flow_field_combined}
\end{figure}

In line with these observations, the results from the C1P3 configuration demonstrate the clear advantages of p-refinement over h-refinement. Despite C1P3 and C2P1 having the same number of degrees of freedom, the use of higher-order polynomial refinement in C1P3 allows for a more precise capture of the flow statistics, particularly in regions close to the wind turbine. 
As illustrated in the flow field visualizations of Figure \ref{fig:flow_field_combined}, C1P3 consistently outperforms C2P1, even with a coarser mesh. This highlights that higher-order polynomial accuracy is more effective at reducing errors than merely increasing h-mesh density. This highlights the efficiency and accuracy benefits of high-order methods and p-refinement when performing actuator line simulations.

\FloatBarrier 

\subsection{Localized p-refinement}

\begin{figure}[h!]
       \begin{center}
       \hspace*{-1.0cm}
\includegraphics[scale=1.0]{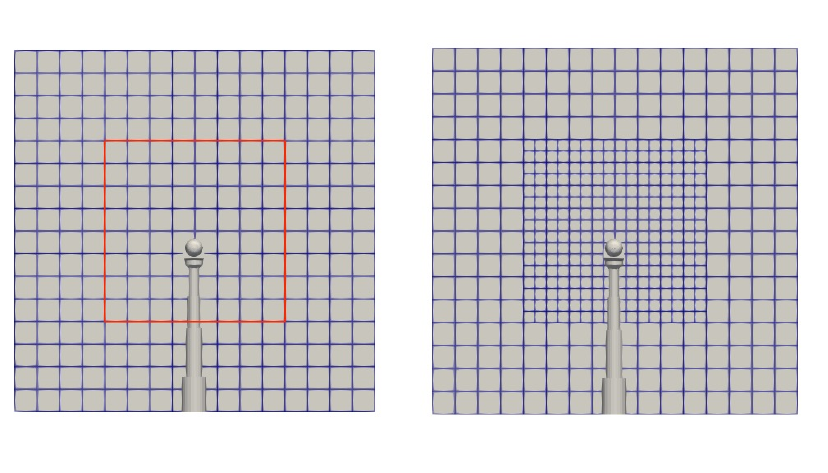}
\end{center}
\caption{Left: y/z-plane of the C1 mesh with the refined region for p-adaption highlighted in red, along with the tower. Right: y/z-plane of the NC mesh with the tower.}
\label{fig:tower_mesh}
\end{figure}

We now examine four setups that use the coarse mesh (C1) with a low polynomial order (P1) outside the wake region and a higher polynomial order (ranging from P2 to P5) within the wake region. These configurations are represented as C1PXP1 in Figures \ref{fig:error1}, \ref{fig:error3}, and \ref{fig:error5}. An inspection of those figures reveals significant stagnation in the error reduction, with minimal improvement observed as the polynomial order in the wake region is increased. This stagnation is likely due to the insufficient resolution in the tower area with C1P1, as the tower is not fully covered by the refined region as seen in Figure \ref{fig:tower_mesh}.

\begin{figure}[h!]
       \begin{center}
\includegraphics[scale=0.6]{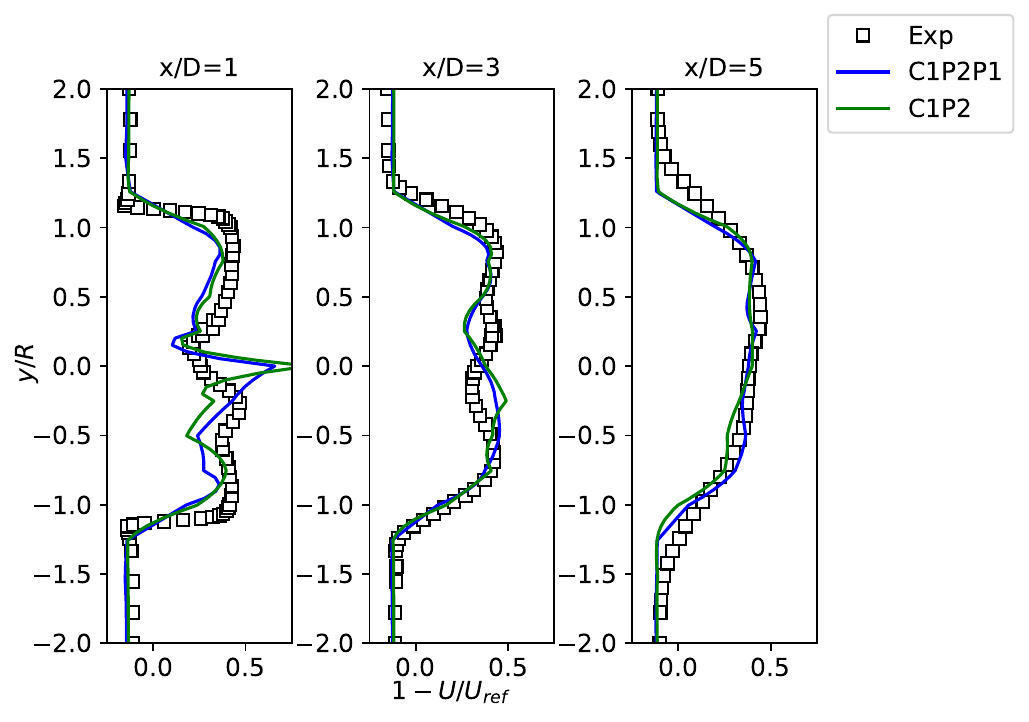}
\end{center}
\caption{Horizontal profiles of mean streamwise velocity deficit at three down-stream positions \begin{math} x/D \end{math} for \begin{math} P=2 \end{math} and adapted ($P=2$ and $P=1$)  with coarse mesh C1 and comparison with experimental data.}
\label{fig:C1P2P1vsC1P2}
\end{figure}

\begin{figure}[h!]
       \begin{center}
\includegraphics[scale=0.6]{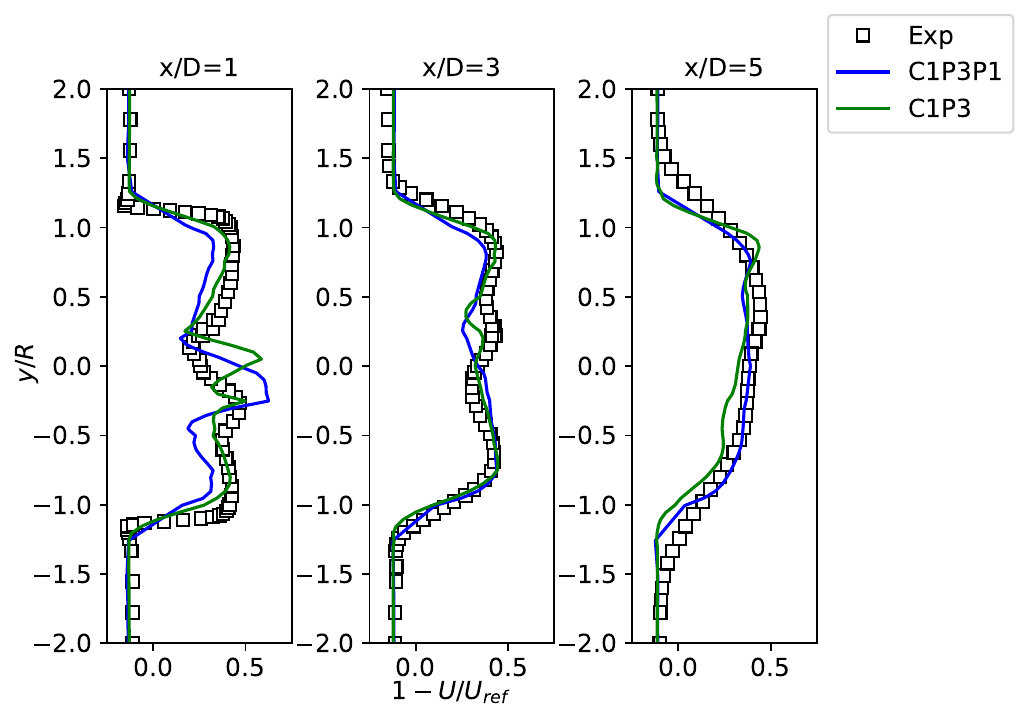}
\end{center}
\caption{Horizontal profiles of mean streamwise velocity deficit at three down-stream positions \begin{math} x/D \end{math} for \begin{math} P=3 \end{math} and adapted ($P=3$ and $P=1$)  with coarse mesh C1 and comparison with experimental data.}
\label{fig:C1P3P1vsC1P3}
\end{figure}

We now compare the setups C1P2P1 versus C1P2, and C1P3P1 versus C1P3 in detail. As shown in Figures \ref{fig:C1P2P1vsC1P2} and \ref{fig:C1P3P1vsC1P3}, the results are quite similar, despite some discrepancies with the tower representation. This observation is also supported by Figures \ref{fig:error1}, \ref{fig:error3} and \ref{fig:error5}  which indicates comparable error levels when using a high polynomial order solely in the wake region.
However, applying a high polynomial order only in the central sections results in a significant reduction in both the number of degrees of freedom and computational cost. For a summary of the degrees of freedom and the associated computational cost reductions achieved by this approach, refer to Table \ref{tab:p-nc-ref} (and Figure \ref{fig:cost}). 

\begin{table}[ht]
 \begin{flushleft}

\begin{tabular}{ |p{1.5cm}||p{3cm}|p{3cm}|p{3cm}|  }
 \hline

 \hline
 Mesh &nDOFs &reduction of nDOFs &reduction of comp. time\\
 \hline

C1P2  & 663,552  &-&-\\
C1P2P1  & 313,344&52.79 \% & 30.99 \%\\
C1P3  & 1,572,864   &-&-\\
C1P3P1  & 540,672&65.63 \% &45.64\%\\

 \hline
\end{tabular}

\caption{Number of degrees of freedom and reduction in computational cost for localized p-refinement. }
 \label{tab:p-nc-ref}
\end{flushleft}
\end{table}
\FloatBarrier

\subsection{Localized h-refinement}

\begin{figure}[h!]
       \begin{center}
\includegraphics[scale=0.6]{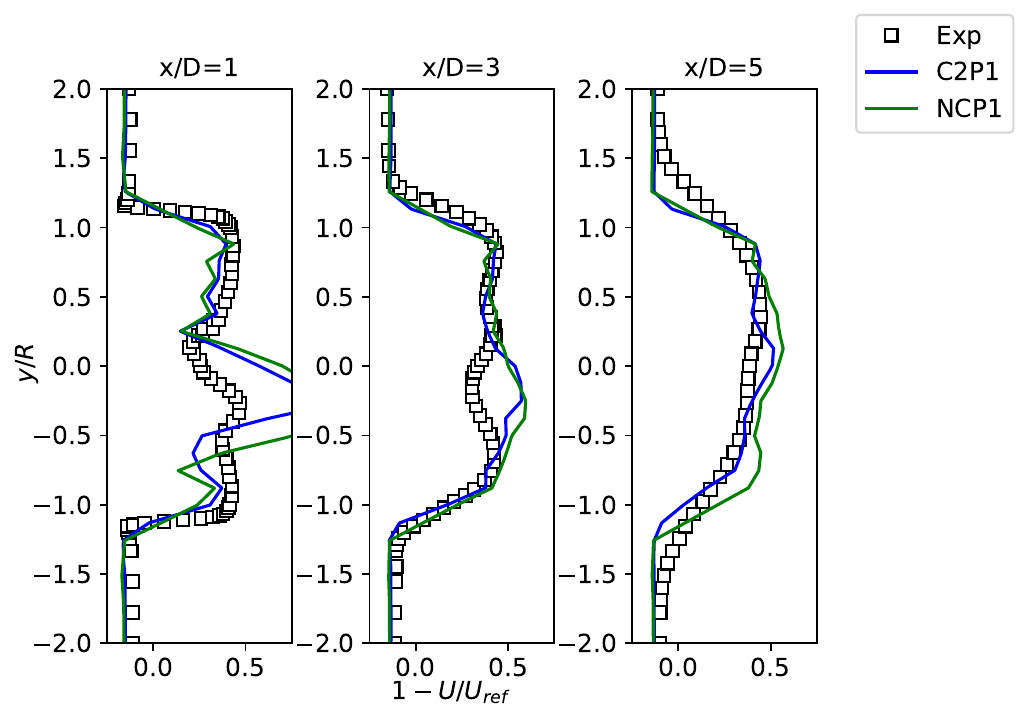}
\end{center}
\caption{Horizontal profiles of mean streamwise velocity deficit at three down-stream positions \begin{math} x/D \end{math} for \begin{math} P=1 \end{math} with fine and non-conforming mesh (C2 and NC) and comparison with experimental data.}
\label{fig:NCP1vsC2P1}
\end{figure}

We now compare NCP1 with C2P1 and NCP2 with C2P2 in detail. The non-conforming (NC) mesh employs C2 resolution in the wake region and C1 resolution outside, which should ideally yield similar error levels with reduced computational cost. Figures \ref{fig:error1}, \ref{fig:error3}, and \ref{fig:error5} indicate that NCP1 has a slightly higher error compared to C2P1, which is expected due to the underresolution in the lower part of the tower. Detailed inspection of the averaged velocity deficit profiles in Figure \ref{fig:NCP1vsC2P1} confirms that the resolution levels are comparable.

\begin{figure}[h!]
       \begin{center}
\includegraphics[scale=0.6]{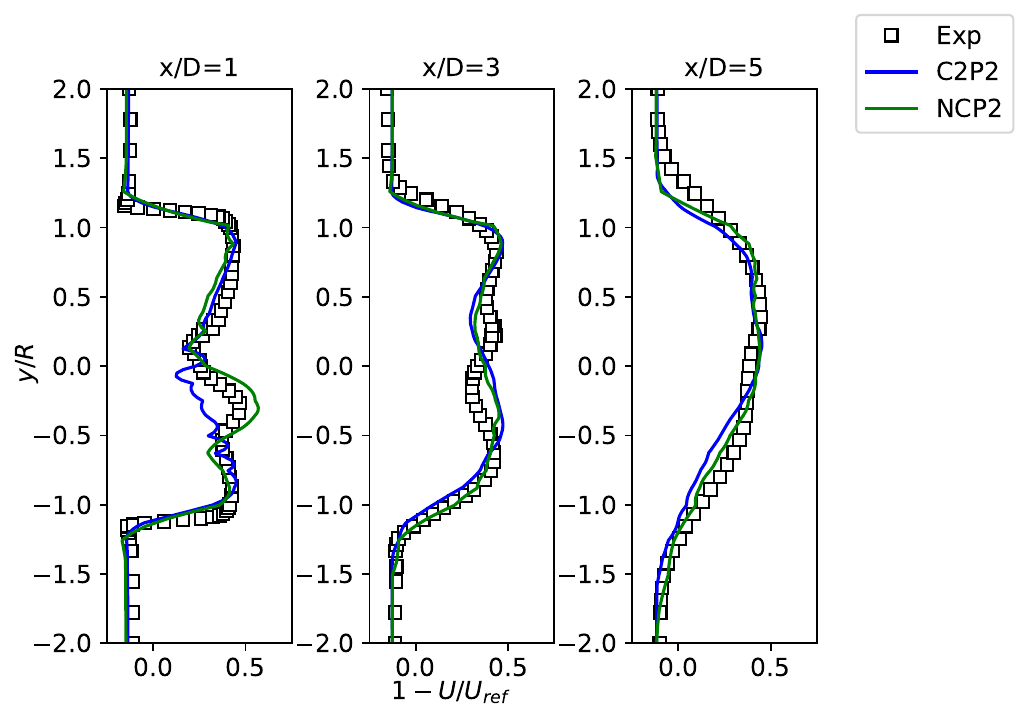}
\end{center}
\caption{Horizontal profiles of mean streamwise velocity deficit at three down-stream positions \begin{math} x/D \end{math} for \begin{math} P=2 \end{math} with fine and non-conforming mesh (C2 and NC) and comparison with experimental data.}
\label{fig:NCP2vsC2P2}
\end{figure}

\begin{figure}[h!]
       \begin{center}
\includegraphics[scale=0.6]{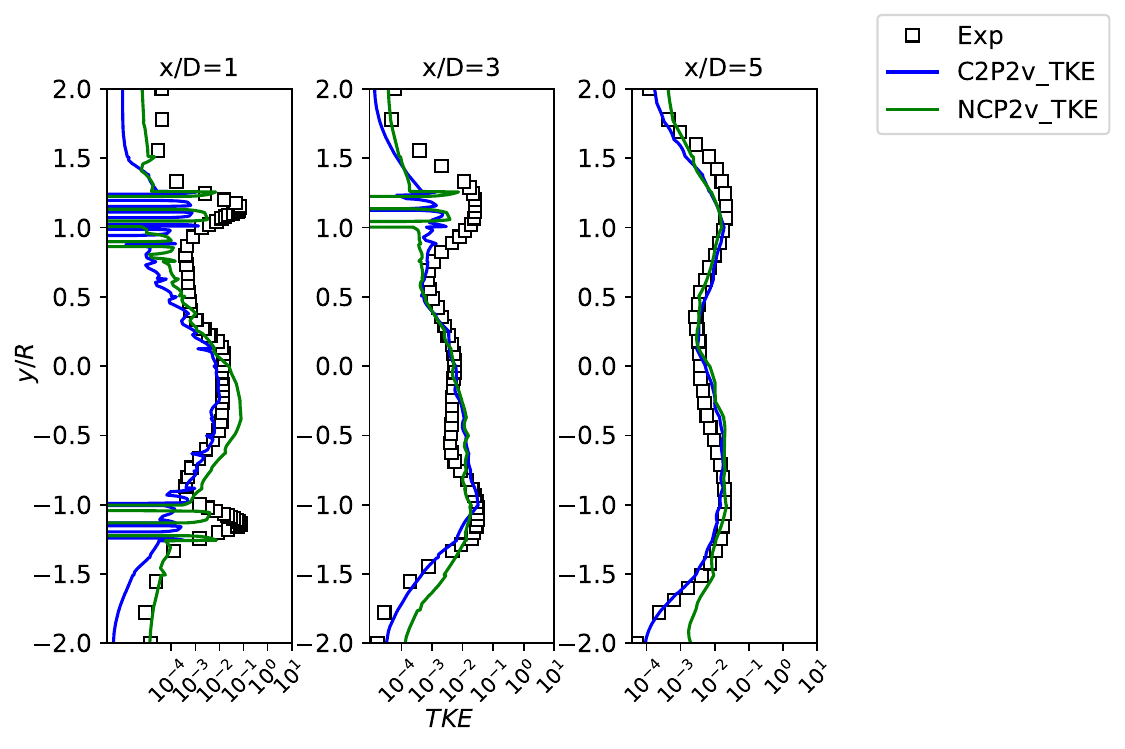}
\end{center}
\caption{Horizontal profiles of the turbulent kinetic energy at three down-stream position \begin{math} x/D \end{math} for \begin{math} P=2 \end{math} with fine and non-conforming mesh (C2 and NC) and comparison with experimental data.}
\label{fig:NCP2vsC2P2tke}
\end{figure}

NCP2 shows error levels comparable to C2P2, and NCP2 even demonstrates slightly lower error at some $x/D$ stations. This can be further examined in Figure \ref{fig:NCP2vsC2P2}, where the velocity deficits are detailed. The unexpected improvement in error may be attributed to a more accurate capture of the solution in the central region at $x/D = 1$. This effect is also seen in other simulations conducted in this study, where a coarse resolution is used outside the wake (e.g., C1P6P1 and C2P3P1). It should be noted that finer meshes used in other studies \cite{marino2024modelling} for the same configuration are closer to C2P2 than to NCP2. Further investigation is required to understand why our solver deviates from the experimental data in this specific region.
To show that the C2P2 solution is indeed superior to the NCP2 solution, Figure \ref{fig:NCP2vsC2P2tke} compares the predictions of turbulent kinetic energy (TKE) of both methods. The C2P2 mesh, which has a higher resolution and more elements than NCP2, demonstrates better alignment with the experimental data. This result is anticipated; a finer, conforming mesh captures turbulent structures more accurately, resulting in TKE predictions that closely match the experimental values. In contrast, while NCP2 remains effective, it shows greater deviations from the experimental data due to its coarser, non-conforming mesh.

As demonstrated in the previous section for the p-refinement, applying the h-refinement to central regions leads to a substantial reduction in both the number of degrees of freedom and computational cost while maintaining similar error levels. For a summary of the degrees of freedom and the corresponding reductions in computational cost achieved by this approach, please refer to Table \ref{tab:h-nc-ref} (and Figure \ref{fig:cost}).
\newline

\clearpage




\begin{table}[ht]
\begin{flushleft}
\hspace*{-1.5cm}
\begin{tabular}{ |p{1.5cm}||p{4cm}|p{4cm}|p{4cm}|  }

 \hline

 \hline
 Mesh &nDOFs &reduction of nDOFs &reduction of comp. time\\
 \hline

C2P1  & 1,572,864  &-&-\\
NCP1  & 540,672  &65.625 \% & 65.77 \%\\
C2P2  & 5,308,416   &-&-\\
NCP2  & 1,824,768  &65.625 \% &64.2\%\\

 \hline
\end{tabular}
\caption{Number of degrees of freedom and reduction in computational cost for localized h-refinement. }
\label{tab:h-nc-ref}
\end{flushleft}
\end{table}
\FloatBarrier




\FloatBarrier
\subsection{The effect of the turbulence model}

\begin{figure}[h!]
       \begin{center}
\includegraphics[scale=0.6]{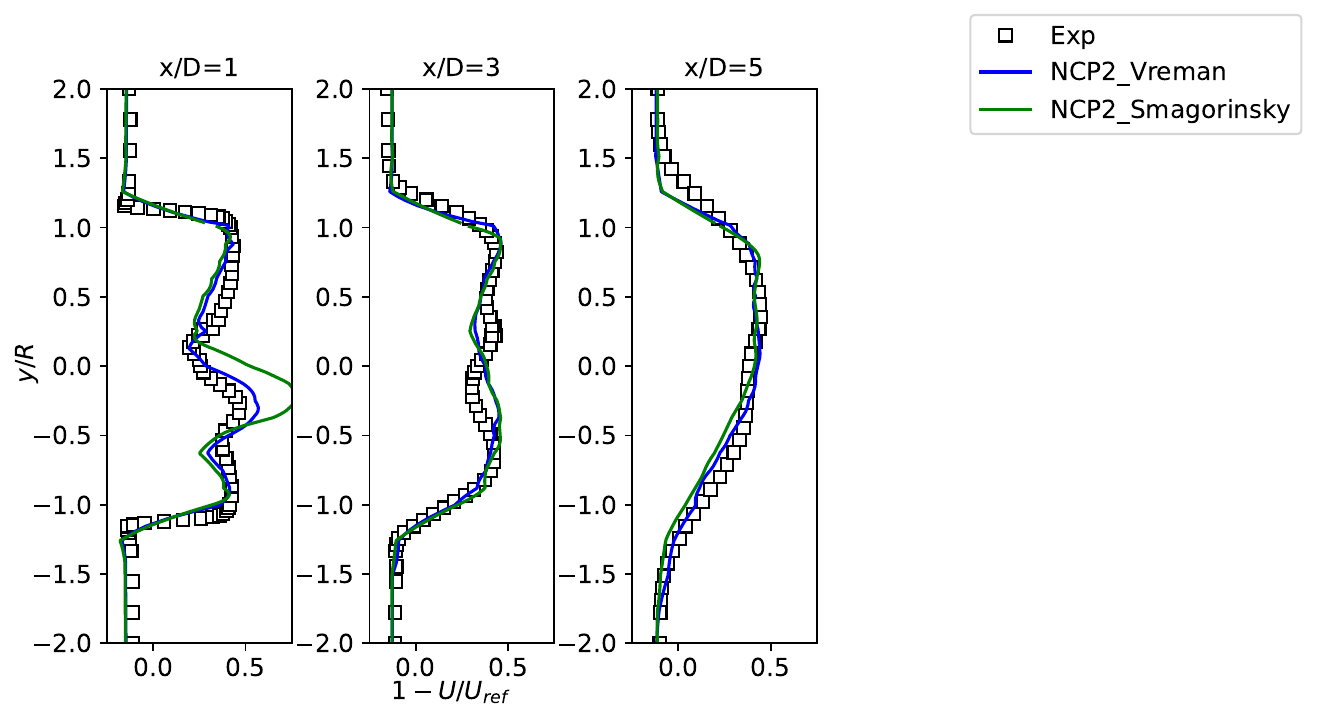}
\end{center}
\caption{Horizontal profiles of mean streamwise velocity deficit at three down-stream positions \begin{math} x/D \end{math} for  \begin{math} P=2 \end{math} with fine mesh and comparison with experimental data.}
\label{fig:NCP2_VvsS}
\end{figure}

\begin{figure}[h!]
       \begin{center}
\includegraphics[scale=0.6]{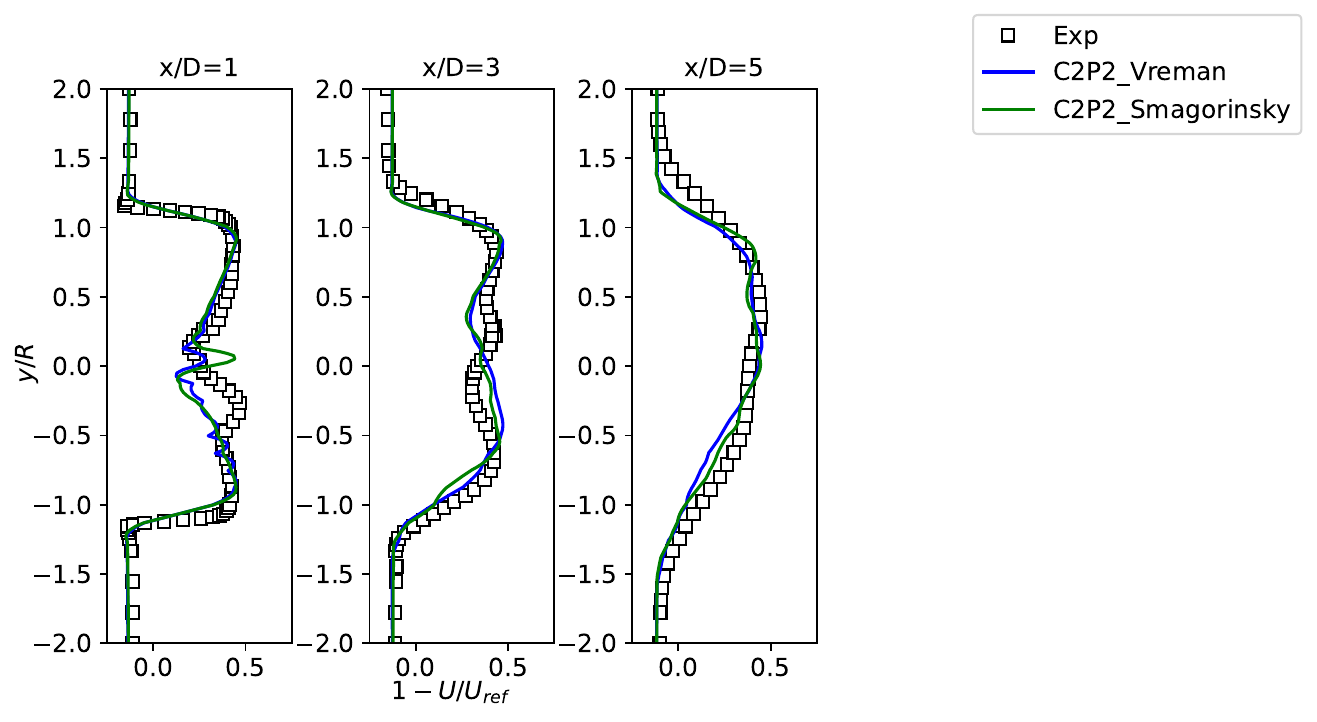}
\end{center}
\caption{Horizontal profiles of mean streamwise velocity deficit at three down-stream positions \begin{math} x/D \end{math} for  \begin{math} P=2 \end{math} with fine mesh and comparison with experimental data.}
\label{fig:C2P2VvsS}
\end{figure}

Throughout this work, we have used the Vreman turbulence model. In this final section, we compare the impact of the turbulence model on the results by comparing the Smagorinsky turbulence model against the Vreman turbulence model for the NCP2 mesh, see Figure \ref{fig:NCP2_VvsS}, and for the C2P2 mesh, see Figure \ref{fig:C2P2VvsS}. Overall, both models provide very similar results, suggesting the limited importance of the model in high-order simulations using actuator lines and immersed boundaries, as presented. Difference are restricted to the hub/nacelle region suggesting some effect of the turbulence mode when interacting with immersed boundaries (and not with the actuator lines). Additionally, we observe the coarser the mesh the more noticeable is the difference between the turbulence models. 

\FloatBarrier

\section{Conclusions}\label{sec:conc}

This research has provided a comparison of the h- and p-refinement methods in the context of high-order methods and wind turbine wake development. Both refinement techniques have shown their strengths and limitations in capturing the intricacies of wind turbine wake, thereby impacting the efficiency of wind farms, which summarized here:
\begin{itemize}
\item The h-refinement method, with its focus on refining the mesh by reducing the size of the elements, has demonstrated its effectiveness in certain scenarios. However, the necessity of hanging nodes presents challenges in computational efficiency.
\item  The p-refinement method, which increases the polynomial degree of the elements, has shown potential in reducing the error exponentially for smooth flows. This suggests its suitability for scenarios where high precision is required.
\end{itemize}
With the implementation of p and h-refinement in our high order solver Horses3d, we obtain faster computations when using p-refinement that h-refinement (for similar number of degrees of freedom) and in addition we have shown that p-refinement can lead to improved accuracy that h-refinement. Overall, we conclude that p-refinement is more beneficial than h-refinement but that the latter should be combined with the former to capture geometrical features such as the turbine hub/nacelle.

The findings of this study contribute to the body of knowledge in wind energy research, particularly in optimizing wind farm operations. The comparative analysis of the h- and p-refinement methods could guide future studies and applications in wind turbine wake modeling. This research underscores the importance of selecting the appropriate numerical method based on the specific requirements and constraints of each scenario in wind energy production. As such, it paves the way for further exploration and innovation in the field of wind energy, with the ultimate goal of enhancing the efficiency and sustainability of wind farms.

\section*{Acknowledgments}
 Hatem Kessasra, aknowledges the Grant 072 Bis/PG/Espagne/2020-2021 of Ministère de l'Enseignement Supérieur et de la Recherche Scientifique, République Algérienne Démocratique et Populaires.
Gonzalo Rubio and Esteban Ferrer acknowledge the funding received by the Grant DeepCFD (Project No. PID2022-137899OB-I00) funded by MICIU/AEI/10.13039/501100011033 and by ERDF, EU.  
Esteban Ferrer would like to thank the support of
Agencia Estatal de Investigación (for the grant "Europa Excelencia 2022" Proyecto EUR2022-134041/AEI/10.13039/501100011033) y del Mecanismo de Recuperación y Resiliencia de la Unión Europea. 

This research has received funding from the European Union (ERC, Off-coustics, project number 101086075).  
Views and opinions expressed here are those of the author(s) only and do not necessarily reflect those of the European Union or the European Research Council. Neither the European Union nor the granting authority can be held responsible for them.
The authors thankfully acknowledge the computer resources at MareNostrum and the technical support provided by the Barcelona Supercomputing Center (RES-IM-2024-2-0001 and IM-2024-1-0003).
Finally, all authors gratefully acknowledge the Universidad Politécnica de Madrid (www.upm.es) for providing computing resources on Magerit Supercomputer.

\appendix
\section{Compressible Navier-Stokes solver}
\label{sec:cNS}
 In this work we solve the 3D Navier-Stokes equations for laminar cases and supplement the equations with the Vreman LES model for turbulent flows. The 3D Navier-Stokes equations when including the  Vreman model can be compactly written as:
%
\begin{equation}
\boldsymbol{q}_t+ \nabla \cdot {\ssvec{F}}_e = \nabla\cdot\ssvec{F}_{v,turb},
\label{eq:compressibleNScompact}
\end{equation}
where $\boldsymbol{q}$ is the state vector of large scale resolved conservative variables $\boldsymbol{u} = [ \rho , \rho v_1 , \rho v_2 , \rho v_3 , \rho e]^T$, $\ssvec{F}_e$ are the inviscid, or Euler fluxes,
\begin{equation}
\ssvec{F}_e = \left[\begin{array}{ccc} \rho v_1 & \rho v_2 & \rho v_3 \\
                                                                                \rho v_1^2 + p & \rho v_1v_2 & \rho v_1v_3 \\
                                                                                	\rho v_1v_2 & \rho v_2^2 + p & \rho v_2v_3 \\
                                                                                	\rho v_1v_3 & \rho v_2v_3 & \rho v_3^2 + p \\
                                                                                	\rho v_1 H & \rho v_2 H & \rho v_3 H
\end{array}\right],
\end{equation}
where $\rho$, $e$, $H=E+p/\rho$, and $p$ are the large scale density, total energy, total enthalpy and pressure, respectively, and $\vec{v}=[v_1,v_2,v_3]^T$ is the large scale resolved velocity components. Additionally, $\ssvec{F}_{v,turb}$ defines the viscous and turbulent fluxes,
\begin{equation}
\ssvec{F}_{v,turb}= \left[\begin{array}{ccc}0 & 0 & 0\\
 \tau_{xx} & \tau_{xy} & \tau_{xz} \\
 \tau_{yx} & \tau_{yy} & \tau_{yz} \\
 \tau_{zx} & \tau_{zy} & \tau_{zz} \\
 \sum_{j=1}^3 v_j\tau_{1j} + \kappa T_x& \sum_{j=1}^3 v_j\tau_{2j} + \kappa T_y& \sum_{j=1}^3 v_j\tau_{3j} + \kappa T_z
\end{array}\right],
\label{eq:viscousfluxes}
\end{equation}
where $\kappa$ is the thermal conductivity, $T_x, T_y$ and $T_z$ denote the temperature gradients and the stress tensor $\boldsymbol{\tau}$ is defined as $\boldsymbol{\tau} = (\mu+\mu_t)(\nabla \vec{v} + (\nabla \vec{v})^T) - 2/3(\mu+\mu_t) \boldsymbol{I}\nabla\cdot\vec{v}$, with $\mu$ the dynamic viscosity, $\mu_t$ the turbulent viscosity (in this work defined through the Vreman 
model) and $\boldsymbol{I}$ the three-dimensional identity matrix. 
The dynamic turbulent viscosity using the Vreman \cite{Vreman_2004} model is given by: 
\begin{equation}
\begin{split}
    &\mu_t = C_v \rho\sqrt{\frac{B_\beta}{\alpha_{ij}\alpha_{ij}}},\\
    &\alpha_{ij} = \frac{\partial v_j}{\partial x_i},\\
    &\beta_{ij} = \Delta^2\alpha_{mi}\alpha_{mj},\\
    &B_\beta = \beta_{11}\beta_{22} -\beta_{12}^2 +\beta_{11}\beta_{33} -\beta_{13}^2 +\beta_{22}\beta_{33} -\beta_{23}^2,
\end{split}
\label{eq-iLES:LES_vreman}
\end{equation}

\noindent where $C_v=0.07$ is the constant of the model. 
The Vreman LES model adjusts the model parameters based on the local flow characteristics and automatically reduces the turbulent viscosity in laminar, transitional, and near-wall regions, allowing to capture the correct physics. 

In the Smagorinsky turbulence model, The dynamic turbulent viscosity using the Smagorinsky model is given by:
\begin{equation}
\mu_t = (\rho C_s \Delta)^2 \sqrt{2S_{ij}S_{ij}}
\end{equation}
where \(C_s=0.2\) is the Smagorinsky constant and \(\Delta=V/(P+1)^3\) is the grid filter defined as the volume of the cell $V$ and scaled by by the polynomial order $P$ (homogeneous in the three directions). The rate-of-strain tensor  \(S_{ij}\) is  defined as:
$
S_{ij} = \frac{1}{2}\left(\frac{\partial v_i}{\partial x_j} + \frac{\partial v_j}{\partial x_i}\right).
$

\section{Mortar element method}
\label{sec:mortar}

In this section, we describe the mortar element method used to couple element faces with different polynomial orders and to connect a single element with multiple elements. The implementation presented follows the approach outlined in \cite{ref118}.

To begin with, the solution values are projected from the faces of the contributing element onto a mortar. In the mortar, the Riemann problem is solved to provide a unique flux, which is then projected back onto the element faces. The projections between the face of the elements and the mortar utilize the least-squares matching of the edge polynomials.
\newline
\begin{figure}[b!]
       \begin{center}
\includegraphics[scale=1.0]{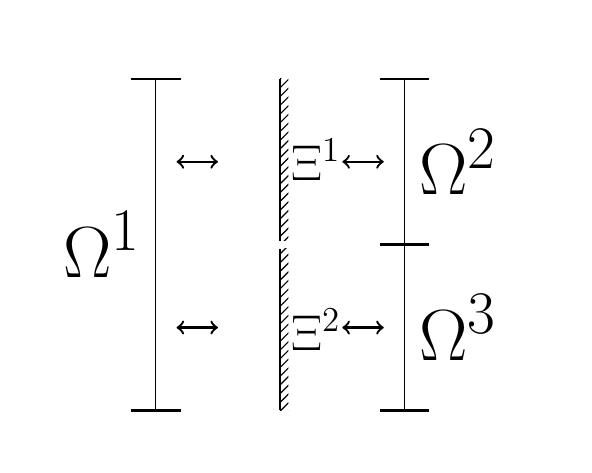}
\end{center}
\caption{Mortar configuration for non conforming interfaces in 2D \cite{ref118}.}
\label{fig:mortar_top}
\end{figure}

Since the system of equations is computed in conservation form, two key conditions must be satisfied at the interfaces. The first is the requirement to maintain global conservation, while the second is termed the 'outflow condition.' Both conditions are met by the least-squares matching of the face and mortar solutions. This least-squares projection ensures that any approximation errors are orthogonal to the polynomial space onto which the solution is projected. In effect, it either truncates or pads the orthogonal polynomial representation of the function, ensuring that the patching introduces no additional errors into modes already represented on the faces of the mortars.

The topology of the mortar subdivision is illustrated in Figure \ref{fig:mortar_top}. With this configuration, the outflow condition can be satisfied \cite{ref28}, as it allows projections that recover the original polynomials across all three elements. This is a distinct advantage compared to the mortar element method for elliptic problems \cite{ref120}, which requires stronger regularity conditions than those for hyperbolic problems \cite{ref118}.

To meet the outflow condition, the order of the mortar polynomials must be sufficiently high. Specifically, the mortar order must be at least as large as the highest subdomain polynomial order among all contributing subdomains. Thus, we choose \begin{math}. J^{1}=max(M^{1},M^{2})\end{math}, \begin{math} J^{2}=max(M^{1},M^{3}) \end{math}
To compute the mortar projection (subdomain to mortar element), we use the unweighted \begin{math} L^{2} \end{math} projection. For each mortar \begin{math} \Xi  \end{math} and each subdomain contributor \begin{math}  \Omega \end{math}, we require that 
\begin{equation}  \int_{-1}^{1} (\Phi(z)-U(o+sz))l^{\Xi } dz =0,\qquad  j=0,1,...,J,   \end{equation}
where $o$ and $s$ are shifts and scaling factors in the integration limits. 
Then the vector of the solution values along the mortar can be computed by
\begin{equation} \Phi=P^{\Omega \rightarrow \Xi }U=M^{-1}SU,  \end{equation} 
where 
\begin{equation}  M_{ij}=\int_{-1}^{1} l_j^{\Xi }l_i^{\Xi } dz         \qquad   i,j=0,1,...,J-1, \end{equation} 
\begin{equation}  S_{ij}^{\Omega}=\int_{-1}^{1} l_j^{\Xi }l_i^{\Omega }(o+sz) dz   \qquad          i=0,1,...,M, j=0,1,...,J. \end{equation} 

For the projection back (mortar element to subdomain), we seek the flux that satisfies
\begin{equation}  \int_{-1}^{1} (F(\xi)-\psi(\xi))l^{\Omega } d\xi =0,\qquad  j=0,1,...,M,   \end{equation}
where 
\begin{equation}
  \psi=\begin{cases}
   \psi^{\Xi^{1}}(\frac{\xi - o^{1}}{s^{1}}),\qquad if \qquad o^{1} \le \xi \le 1,\\
    \psi^{\Xi^{2}}(\frac{\xi - o^{2}}{s^{2}}),\qquad if \qquad 0 \le \xi \le o^{1}.
  \end{cases}
\end{equation}
If we now define
\begin{equation}  M_{ij}=\int_{-1}^{1} l_j^{\Omega }l_i^{\Omega } d\xi         \qquad   i,j=0,1,...,M, \end{equation} 
\begin{equation}  S_{ij}^{k}=s^{k}\int_{-1}^{1}l_i^{\Xi}l_j^{k}(o^{k}+s^{k}z) dz   \qquad          i=0,1,...,J, j=0,1,...,M, \end{equation} 
we can write 
\begin{equation} F^{1}=\sum_{k=1}^{2} P^{k}\psi^{k},
\end{equation}
where \begin{math} P^{k}=P^{\Xi \rightarrow \Omega^{k}}=M^{-1}S^{k} \end{math} is the projection matrix
\newline
The use of the unweighted \begin{math} L^{2} \end{math} projections gives the mortar approximation two desired properties: global conservativeness and satisfaction of the outflow condition. In the case of a conforming interface (i.e., no hanging nodes on the interface, i.e., \begin{math} o=0 \end{math} and \begin{math} s=1 \end{math}), if the mortar order and the subdomain polynomial order are equal, the projection simplifies to the identity. Consequently, a simple copy of the flux from the mortar to the face can be made.

\begin{figure}[h!]
       \begin{center}
\includegraphics[scale=1.5]{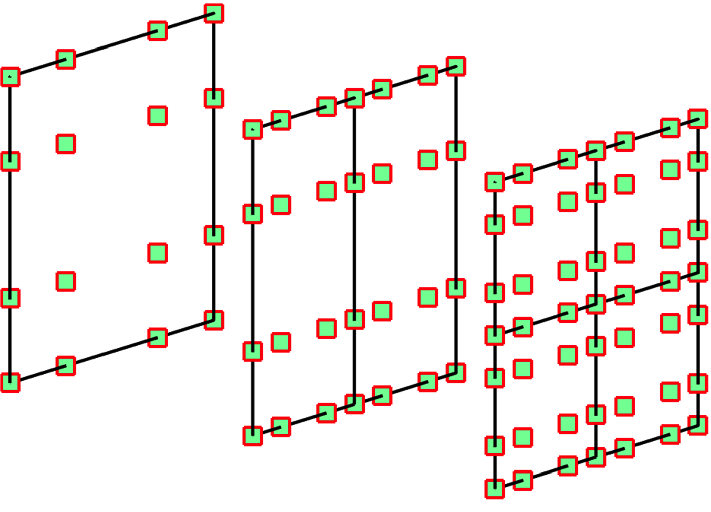}
\end{center}
\caption{Mortar configuration for non conforming interfaces in 3D.}
\label{fig:mortar-3d}
\end{figure}

In the 3D formulation, the same 1D matrices are used for the mortar projection, which is carried out in two stages: first along the vertical direction, then along the horizontal direction, as shown in Figure \ref{fig:mortar-3d}.

\section{Summary of identified sources of modeling errors}
\label{sec:modelingerrors}
Modeling errors are the reason why numerical simulations may not fully match reality, even with infinitely fine mesh resolution. In this appendix, we summarize the key sources of modeling errors identified in our approach:
\begin{itemize}
     \item \textbf{Inlet turbulence}: No turbulence is considered at the inlet in our simulations.

    \item \textbf{Tower model}: The tower is modeled using an Immersed Boundary Method (IBM), which yields lower accuracy compared to a body-fitted mesh.

    \item \textbf{Turbulence model}: We perform Large Eddy Simulations (LES) using the Vreman turbulence model. While highly accurate, Direct Numerical Simulation (DNS) would provide even greater precision.

    \item \textbf{Wind tunnel walls}: To reduce computational costs, free-slip wall conditions are applied. However, using no-slip walls and applying mesh refinement near the walls could improve simulation accuracy.

    \item \textbf{Wind tunnel blockage}: There is a mismatch between the dimensions of the wind tunnel and our simulation domain, which is $0.35$~m shorter and $0.91$~m narrower. This discrepancy may cause blockage effects and acceleration in the simulation, which could be mitigated by using a larger mesh.

    \item \textbf{Wind turbine blades}: We model the wind turbine blades using an actuator line approach. Greater accuracy could be achieved with sliding-body-fitted meshes.

\end{itemize}

Addressing these issues would lead to more accurate simulations, but at the cost of increased computational resources.


 \bibliographystyle{elsarticle-num} 
 \bibliography{bibrefs}

\end{document}